# Jahn-Teller states mixed by spin-orbit coupling in an electromagnetic field

A. S. Miñarro, G. Herranz
*Institut de Ciència de Materials de Barcelona (ICMAB-CSIC), Campus UAB, 08193 Bellaterra, Catalonia, Spain*

Spin-orbit coupling plays a pivotal role in condensed matter physics. For instance, spin-orbit interactions affect the magnetization and transport dynamics in solids, while spins and momenta are locked in topological matter. Alternatively, spin-orbit entanglement may play an important role in exotic phenomena, like quantum spin liquids in 4d and 5d systems. An interesting question is how electronic states mixed by spin orbit coupling interact with electromagnetic fields, which may hold potential to tune their properties and reveal interesting physics. Motivated by our recent discovery of large gyrotropic signals in some Jahn-Teller manganites, here we explore the interaction of light with spin-mixed $t_{2g} - e_g$ states in a $d^4$ transition metal. We show that spin-orbit mixing enables electronic transitions that are sensitive to circularly polarized light, giving rise to a gyrotropic response that increases with spin-orbit coupling. Interestingly, photoexcited transitions that involve spin reversal are behind such gyrotropic resonances. Additionally, we find that the interaction with the electromagnetic field depends strongly on the relative orientation of the propagation of light with respect to Jahn-Teller distortions and spin quantization. We suggest that such interactions offer the opportunity to use electromagnetic waves at optical wavelengths to entangle orbital and spin degrees of freedom. Our approach, which includes a group-theoretical treatment of spin-orbit coupling, has wide applicability and provides a versatile tool to explore the interaction of electromagnetic fields with electronic states in transition metals with arbitrary spin-orbit coupling strength and point-group symmetries.

## I. INTRODUCTION

Orbital degrees of freedom are an essential ingredient of the physics and chemistry of transition metal compounds [2–5]. The coupling of orbitals to spin, charge or lattice determines many properties of solids and molecules. In the presence of orbital degeneracy, the symmetry of non-linear molecules and solid state systems is broken spontaneously through the Jahn-Teller effect [6–8]. This phenomenon has far-reaching consequences in spectroscopic and chemical properties [9–11], and is also responsible for the emergence of nontrivial quantum effects, associated with the appearance of rotational quantization of vibronic states and geometric phases [12–14]. On the other hand, spin orbit interactions are key to new developments in classical and quantum computation and lie beneath new discoveries in condensed matter physics related to topological matter, such as the quantum spin Hall effect or the realization of topological insulators and Weyl semimetals [15–17] and Kitaev physics in quantum spin-liquids [18–21]. At the same time, there is a highly nontrivial interplay between spin-orbit coupling and the Jahn-Teller effect when $t_{2g}$ states are partially filled, where entangled quantum spin-orbital states may emerge [22–24].

An interesting question is how electromagnetic fields interact with spin-orbit mixed states, which could pave the ground to explore quantum physics in these systems. Motivated by our recent finding of large gyrotropic signals in $La_{2/3}Ca_{1/3}MnO_3$ (originated by the different optical response to light of opposite handedness in the presence of Jahn-Teller distortions) [25], we present here a group-theoretical analysis to study this problem. Our formalism has general applicability and provides a useful route to extend the analysis to heavy transition metals in arbitrary point-group symmetries. In the following, we describe in great detail the interaction with an electromagnetic field of spin-orbit mixed $t_{2g} - e_g$ states in a $3d$ metal, which provides the clues to its generalization to other transition metals.

We first note that when dealing with spin-orbit physics in light transition metals, the mixing between $t_{2g}$ and $e_g$ orbitals is usually neglected, since crystal-field splitting and exchange energies are much larger than spin-orbit coupling [22]. However, this approximation breaks down under particular conditions. To illustrate this point, we consider the Tanabe-Sugano diagram for the case of an ion with $d^4$ configuration in $O_h$ symmetry [1, 26]. We see that for values of the crystal field $10Dq$ and the Racah parameter $B$ that fulfill the condition $(Dq/B)_c \leq 2.7$ the ground state term is $^5E_g$, whereas for large enough $Dq/B$ the ground state is $^3T_{1g}$ (Fig. 1a). In both limits, a good approximation is that spin-orbit interactions act only on the $t_{2g}$ manifold (the orbital moment is quenched for $e_g$ states), and the spin-orbit mixing of $e_g$ and $t_{2g}$ states (and, therefore, between $^5E_g$ and $^3T_{1g}$) can be ignored, at least to first order in spin-orbit coupling. This results in the so-called $T-P$ equivalence, where the spin-orbit physics of the $t_{2g}$ manifold can be described with an effective angular momentum $L = 1$, like for $p$ orbitals [1].

However, this approximation breaks down for specific situations. For instance, in $4d/5d$ transition metals, where several energy scales (including spin-orbit coupling and crystal-field) are comparable [27, 28], the $t_{2g} - e_g$ mixing can not be generally ignored [29, 30], which is relevant to predict magnetic excitations in heavy transition metal oxides. Alternatively, the $T - P$ equivalence fails when the difference in energies between spectroscopic terms becomes comparable with spin-orbital coupling. This may happen for ratios $(Dq/B)_c \approx 2.7$ in the Tanabe-Sugano diagram [1], which may occur in $3d$-transition metals (Fig. 1(a)). Alternatively, a reduction from $O_h$ to $D_{4h}$ symmetry (e.g., induced by a Jahn-Teller instability, as we discuss below), may lead to spin-orbital mixing between spectroscopic terms $^5A_{1g}$ and $^3E_g$ away from $(Dq/B)_c \approx 2.7$, see Fig. 1(b). As we show below, spin-orbital mixing between $t_{2g}$ and $e_g$ states –which may be also induced by electronic correlations or structural distortions– enables optical transitions that can be probed by circularly polarized light. The sensitivity to circular polarization stems from terms of different spin multiplicity mixed by spin-orbit coupling (e.g., $S$ =1 for $^3E_g$ and $S$ =2 for $^5A_{1g}$ in $D_{4h}$ point symmetry), which allows optical transitions between states with different spin projections, which, otherwise, are absent in the absence of spin-orbit mixing. For the sake of conciseness, we restrict our discussion to $3d^4$ ions in which spin-orbital mixing is induced by symmetry reduction to $D_{4h}$ induced by Jahn-Teller instabilities (Fig. 1b). At the end we discuss briefly how the group-theoretical analysis can be extended to other transition metals with arbitrary spin-orbit coupling strength and their interaction with electromagnetic radiation.

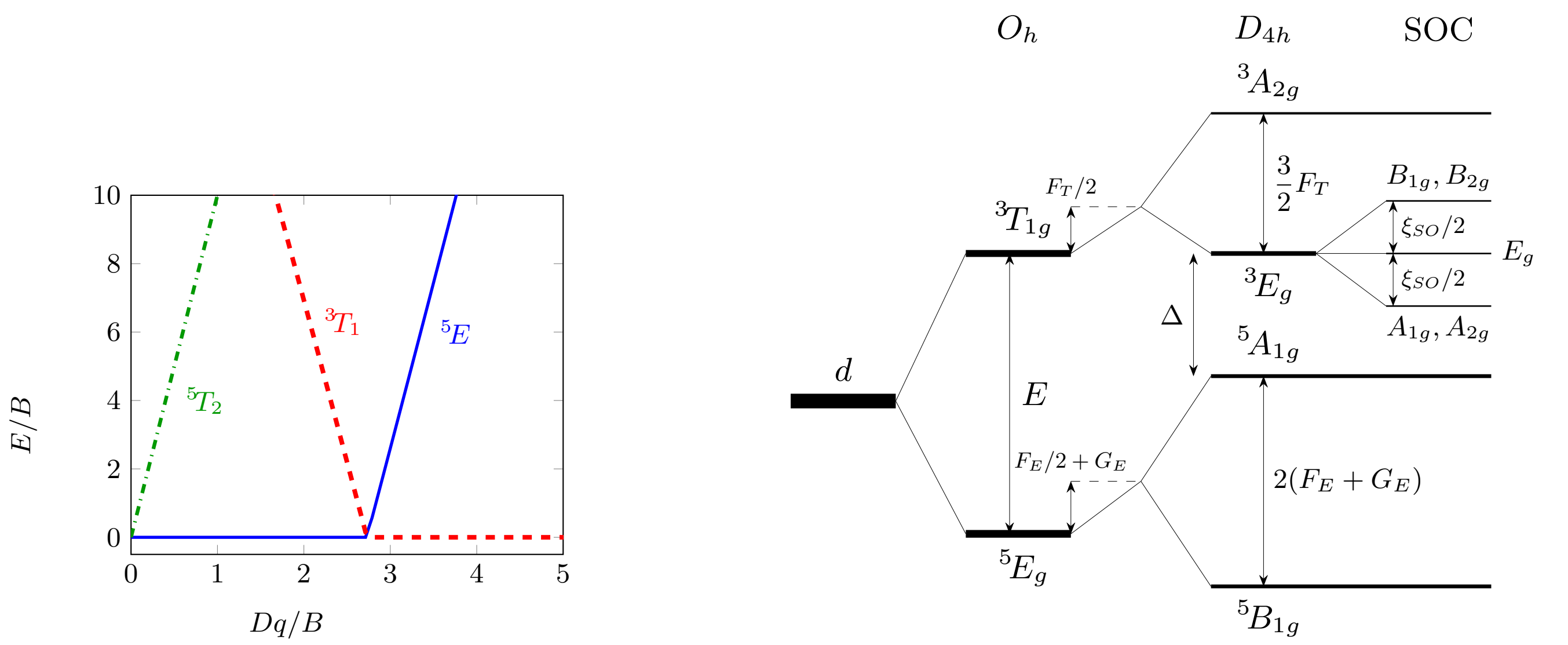

(a) (b)

FIG. 1. a) Tanabe-Sugano diagram for a $d^4$ configuration, representing the lowest energy terms $^5T_2$, $^3T_1$ and $^5E$ in $O_h$ point symmetry [1]. Along the ordinate axis, the energy $E$ is shown relative to the Racah parameter $B$. The abscissa displays the ratio between the "differential of quanta" $Dq$ and $B$ (for octahedral complexes the crystal field energy is $10Dq$). b) Splitting of the spectroscopic lines in $O_h$, $D_{4h}$ and spin-orbit coupling (SOC) point symmetries. In this work, we consider that the reduction to $D_{4h}$ symmetry is driven by Jahn-Teller interactions.

To describe the physics of spin-orbital mixing we consider a Hamiltonian that has the following form:

$$\mathcal{H} = \mathcal{H}_S + \mathcal{W} = \sum_i \Bigg\{ \sum_{\psi_i} E_\psi |\psi_i\rangle\langle\psi_i| + \sum_{\psi_i,\phi_i} V_{\psi\phi}|\psi_i\rangle\langle\phi_i| \Bigg\} + \sum_{\substack{i\neq j \\ \psi_i,\psi_j}} \alpha_{ij}|\psi_i\rangle\langle\psi_j| \tag{1}$$

where $\mathcal{H}_S$ stands for on-site interactions and $\mathcal{W}$ represents the interaction of electrons with an electromagnetic field. We consider that the interaction with light induces the hopping with amplitude $\alpha_{ij}$ of the fourth electron in the $d^4$ ion to neighboring $d^3$ sites $i$, $j$ in the lattice. The on-site Hamiltonian $\mathcal{H}_S$ contains diagonal terms denoted by $E_\psi|\psi_i\rangle\langle\psi_i|$ and off-diagonal $V_{\psi\phi}|\psi_i\rangle\langle\phi_i|$ contributions, coming from vibronic couplings and spin-orbit interactions. The Hamiltonian can be formally expressed in terms of irreducible representations $|\psi_i\rangle, |\phi_i\rangle \in \{\ ^3A_{2g},\ [B_{1g}+B_{2g}],\ E_g,\ [A_{1g}+A_{2g}],\ ^5A_{1g},\ ^5B_{1g}\}$. The terms of this basis are expressed as linear combinations of Slater determinants of monoelectronic orbitals $t_{2g} \in (|xy\rangle, |yz\rangle, |xz\rangle)$ and $e_g \in (|x^2-y^2\rangle, |z^2\rangle)$ that respect the Pauli exclusion principle and the point-group symmetries in orbital ($D_{4h}$, due to Jahn-Teller instabilities) and spin spaces (in Appendix A we give a detailed derivation of this basis and the development of the corresponding Slater determinants). We note that $(B_{1g}, B_{2g})$ and $(A_{1g}, A_{2g})$ are degenerate (possibly this accidental degeneracy is broken if we consider higher-order relativistic corrections) and they are lumped together in the ba-sis $|\psi_i\rangle$. Additionally, the presence of magnetic fields can lift the $2S+1$-fold degeneracy of terms $^3A_{2g}$ $(S=1)$, $^5A_{1g}(S=2)$ and $^5B_{1g}$ $(S=2)$, giving a total of 19 states for the full dimensionality of the $|\psi_i\rangle$ basis. The on-site Hamiltonian $\mathcal{H}_S$ can be decomposed as:

$$\mathcal{H}_S = \mathcal{H}_0 + \mathcal{H}_{JT} + \mathcal{H}_{SO} \tag{2}$$

where $\mathcal{H}_0$ includes the energy splitting between $O_h$ point-group terms $^5E_g$ and $^3T_{1g}$ due to crystal field and exchange interactions (see Fig. 1b), while $\mathcal{H}_{JT}$ takes into account interactions with Jahn-Teller modes and $\mathcal{H}_{SO}$ is the spin-orbit coupling contribution.

The paper is organized as follows. In Sec. II A, we describe the spontaneous breaking of orbital degeneracy driven by Jahn-Teller instabilities in a $d^4$ ion under $O_h$ symmetry. In this situation, $^5E_{2g}$ and $^3T_{1g}$ electronic states interact with doubly degenerate $E_g$ Jahn-Teller modes, resulting in $E\otimes e$ and $T\otimes e$ vibronic interactions. As a result, the point-group symmetry is reduced to $D_{4h}$ and the terms split into $^3A_{2g}$, $^3E_g$, $^5A_{1g}$ and $^5B_{1g}$ states (see Fig. 1b). In Sec. II B we study the point symmetries in orbital and spin spaces related to the spin-orbit operator. The combination of Jahn-Teller and spin-orbit interactions split further these terms into irreducible representations $|\psi_i\rangle \in \{\,^3A_{2g}, [B_{1g}+B_{2g}], E_g, [A_{1g}+A_{2g}], \,^5A_{1g}, \,^5B_{1g}\}$, see Fig. 1b and Appendix A. In Sec. III, we describe the interaction of electrons with electromagnetic fields. As a result of this interaction, we assume that electrons hop between neighboring sites. We focus on the problem of a lattice in which the fourth electron of an isolated Jahn-Teller $d^4$ ion hops to $d^3$ nearest-neighbours. We assume that the solid is a transition metal oxide with perovskite structure with $ABO_3$ chemical formulation, in which generally $A$ is a rare-earth element, $B$ a transition metal and $O$ is oxygen. These systems form a large family of materials that includes $La_{2/3}Ca_{1/3}MnO_3$, with a broad diversity of physical properties, including magnetism, ferroelectricity or superconductivity [31–34]. The undisturbed perovskite is formed by octahedral unit cells with the metal $B$ sitting at the center of an octahedron formed by six ligand oxygen anions with $O_h$ point symmetry [35, 36], see Fig. 2(a). Using the formalism of two-center Slater-Koster integrals, we derive analytic expressions for the light-induced hopping amplitudes between lattice sites. The perturbative analysis discussed in Sec. IV demonstrates that spin-orbit coupling and intraatomic $t_{2g}-e_g$ mixing are essential to the appearance of gyrotropic responses, and that the latter involve photoexcitations in which one of the spins is inverted. Remarkably, this observation opens the possibility of using electromagnetic fields to manipulate spins via the mechanism described here. Subsequently, in Sec. V, we analyze the electronic response to circularly polarized electromagnetic waves. For that purpose, we analyze the density of $^5B_{1g}$ states from the imaginary part of quantum propagators of the different electronic orbitals and obtain expressions for their spectral functions. In Sec. VI we analyze these spectral functions in circularly polarized electromagnetic fields as a function of the relative strength of Jahn-Teller and spin-orbit interactions. From this analysis, we extract information about the gyrotropic responses, by which the polarization of light is changed as a result of the interactions with $t_{2g}-e_g$ spin-orbit mixed states. Finally, in Sec. VII, we summarize the main results and discuss perspectives of further work, especially the possibility of entangling spin and orbital degrees of freedom using electromagnetic waves, which could be relevant in the framework of non-trivial quantum states in other systems, including heavy $4d$-$5d$ transition metals.

## II. THEORY: ON-SITE INTERACTIONS

### A. Jahn-Teller interactions

We first derive the Hamiltonian terms for the interaction of electron orbitals with Jahn-Teller modes. Under $O_h$ point-group symmetry they can interact with two degenerate representations for $E_g$ Jahn-Teller vibrational modes corresponding to tetragonal modes $Q_{E_g u} = Q_3 = 2\Delta z - \Delta x - \Delta y$ and orthorhombic modes $Q_{E_g v} = Q_2 = \sqrt{3}(\Delta x - \Delta y)$, respectively (see Fig. 2). We thus need to solve the $E\otimes e$ and $T\otimes e$ problems to derive analytic expressions for the corresponding vibronic interactions between Jahn-Teller $E_g$ modes and doubly degenerate $E_g$ and triply degenerate $T_{1g}$ electronic states in $O_h$ symmetry [37]. A convenient way to derive these expressions is to write the Jahn-Teller modes in terms of an angle $\vartheta$ as $Q_3 = \cos\vartheta$ and $Q_2 = \sin\vartheta$ [38, 39]. Using Pauli matrices $\upsilon_i$ in the pseudospin space of $[\,^5A_{1g}, \,^5B_{1g}]$ states in $D_{4h}$ symmetry, the $E\otimes e$ Jahn-Teller interaction can be expressed as:

$$\begin{aligned}\mathcal{H}_{JT}^{E\otimes e} = \frac{F_E + 2G_E}{2}\upsilon_0 + (F_E + G_E)\upsilon_z + \\ + (F_E - 2G_E)\delta\vartheta\upsilon_x\end{aligned} \tag{3}$$

where $F_E$ and $G_E$ are linear and quadratic vibronic constants, and $\delta\vartheta$ represents perturbative $Q_2$ orthorhombic distortions that will be described below. The dependence of Jahn-Teller modes on $\vartheta$ defines a potential energy surface, which, in the case of harmonic approximation (i.e., $G_E = 0$), defines a "Mexican hat" [40]. In solids, however, anharmonic contributions are usually relevant and quadratic constants (so that $G_E \neq 0$) must be included in $Eq.$(3). As a result, the surface potential warps producing three minima at $\vartheta_n = 2n\pi/3$, which correspond to tetragonal elongations of the octahedral cell along $z$, $y$ and $x$ axes for $n = 0, 2, 1$, which stabilize the occupation of $d_{z^2}$, $d_{y^2}$ and $d_{x^2}$ orbitals, respectively [41–43] (see also Appendix B for a detailed description of the Jahn-Teller Hamiltonian and the vibronic interactions). Actually, the stabilization of tetragonal Jahn-Teller distortions in solids has been confirmed experimentally in a large number of compounds [44], including manganites [45–48].

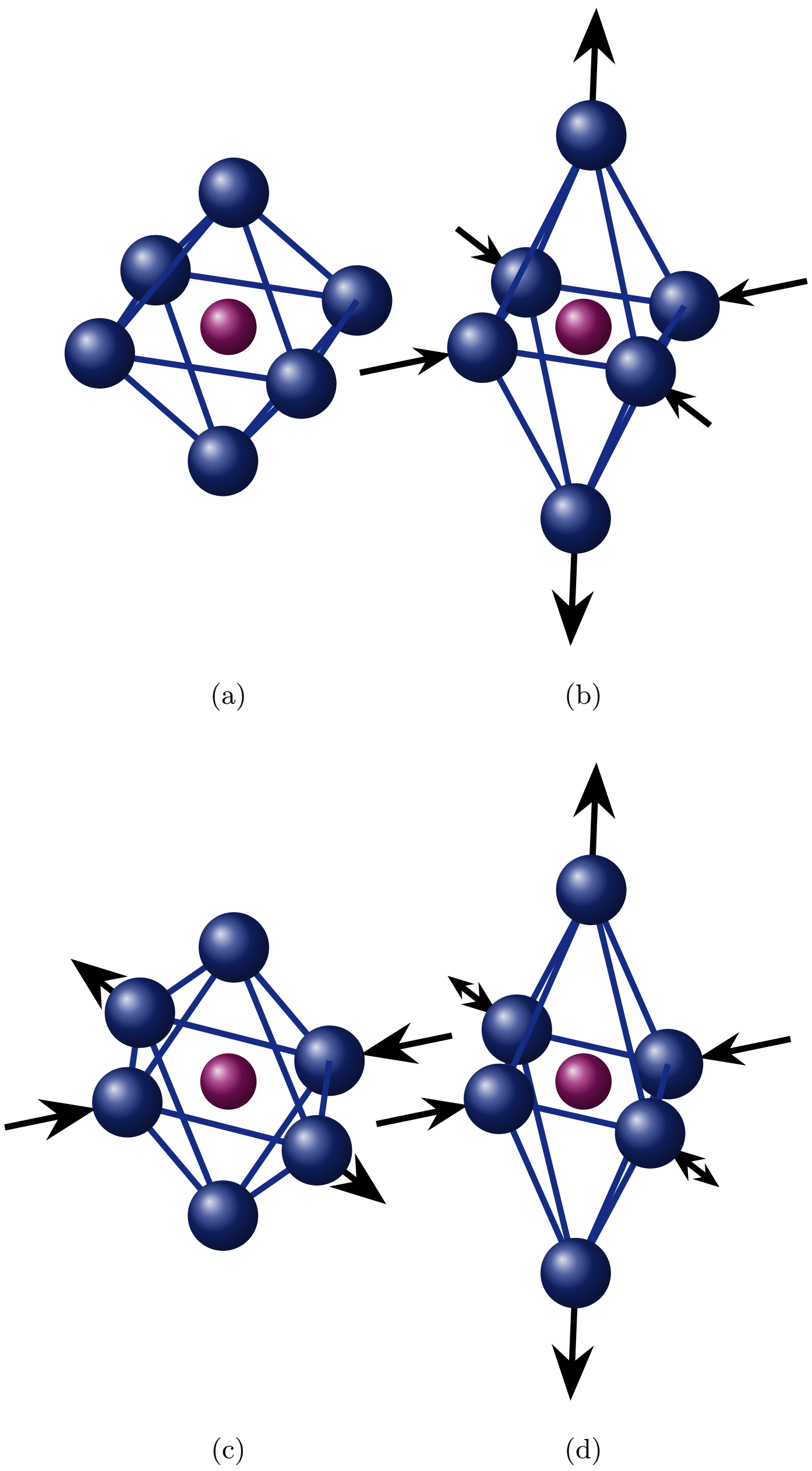

FIG. 2. a) Depiction of a transition metal in an octahedral coordination with six oxygen ligands. b) Tetragonal Jahn-Teller distortion corresponding to the mode $Q = Q_3$. c) Orthorhombic Jahn-Teller distortion corresponding to the mode $Q = Q_2$. d) Superposition of the two distortion modes corresponding to $Q = Q_3 \cos\vartheta + Q_2 \sin\vartheta$.

Now we derive the analytic expressions for vibronic interactions involving $T_{1g}$ electronic states, which can interact with $E_g$ and $T_{2g}$ representations of Jahn-Teller modes [37]. Consequently, such derivation requires solving the $T \otimes e$ and $T \otimes t$ problems [41]. We assume, however, that the $T \otimes t$ contribution is negligible, since in our case lattice deformations are predominantly driven by Jahn-Teller instabilities of electrons in $e_g$ states. Therefore, we only consider the contribution of $T \otimes e$ to the Hamiltonian as follows:

$$\mathcal{H}_{JT}^{T\otimes e} = \frac{1}{2} F_T \left[ \lambda_0 - \sqrt{3}\lambda_8 - \delta\vartheta\lambda_3 \right] \tag{4}$$

where $F_T$ is the vibronic coupling constant for $T \otimes e$ and $\lambda_2$ and $\lambda_8$ are Gell-Mann matrices, defined as

$$\lambda_3 = \begin{pmatrix} 1 & 0 & 0 \\ 0 & -1 & 0 \\ 0 & 0 & 0 \end{pmatrix} \tag{5a}$$

$$\lambda_8 = \frac{1}{\sqrt{3}} \begin{pmatrix} 1 & 0 & 0 \\ 0 & 1 & 0 \\ 0 & 0 & -2 \end{pmatrix} \tag{5b}$$

As mentioned above, anharmonic lattice contributions stabilize tetragonal elongations along the three main axes of the $ABO_3$ octahedral cell units, denoted as $\vartheta_n = 2n\pi/3$. This entails breaking the degeneracy of ${}^5E_{2g}$ electronic states in $O_h$ into ${}^5A_{1g}$ and ${}^5B_{1g}$ terms in $D_{4h}$ symmetry, being the latter lower in energy for the elongated tetragonal distortions (Fig. 1b). In this work we also consider small orthorhombic Jahn-Teller distortions corresponding to $Q_2$ modes, which modify perturbatively the elongated tetragonal distortions, so that the angle in $Q_2 - Q_3$ space is mapped to $\vartheta_n \rightarrow \vartheta_n + \delta\vartheta_n$. As can be inferred from the $E \otimes e$ Hamiltonian ((2)), these orthorhombic perturbations induce non-diagonal transitions between ${}^5A_{1g}$ and ${}^5B_{1g}$ terms. On the other hand, for the $t_{2g}$ sector, the reduction to $D_{4h}$ symmetry splits the ${}^3T_{1g}$ term into single-degenerate ${}^3A_{2g}$ and doubly degenerate ${}^3E_g$ terms, being the latter lower in energy. In this case, it can be shown that, to first-order, $\vartheta_n + \delta\vartheta_n$ perturbations do not mix ${}^3A_{2g}$ and ${}^3E_g$ terms.

### B. Spin-orbit coupling

In order to compute the matrix elements of spin-orbit coupling in $D_{4h}$ symmetry we use the operator equivalent method [1, 49]. In this approach, the spin-orbit operator is defined by linear combinations $V_{\lambda q}^{\Lambda} = T_{\lambda}^{\Lambda} S_q^1$, where $T_{\lambda}^{\Lambda}$ corresponds to irreducible representations $\Lambda$ in the orbital space with basis $\lambda$ and $S_q^1$ corresponds to irreducible representations $D_q^{(1)}$ in the spin-rotation group. Since the $T \otimes e$ Hamiltonian term in Equation (3) is relatively small –i.e., $F_T \delta\vartheta \lessapprox \xi_{SO}$, where $\xi_{SO}$ is the spin-orbit coupling constant– we can study the orbital space $T_{\lambda}^{\Lambda}$ in the $D_{4h}$ point-group. Therefore, while $T_{\lambda}^{\Lambda}$ transforms as $T_{1g}$ in $O_h$ symmetry, the reduction to $D_{4h}$ implies that $T_{\lambda}^{\Lambda}$ transforms according to irreducible representations $A_{2g}$ with spatial symmetry $\nu$ and $E_g$ with spatial symmetries $\kappa, \mu$ (see Appendix A 2 b for a definition of these symmetries). On the other hand, the spin part $S_q^1$ is expressed using spherical coordinates, with quantum numbers $q = 0, \pm 1$. Taking this into account, it can be demonstrated that the spin-orbit coupling Hamiltonian can be expressed in terms of operator equivalent matrices $V_{\kappa\pm1}^{E_g}$, $V_{\mu\pm1}^{E_g}$ and $V_{\nu0}^{A_{2g}}$ as follows:

$$\mathcal{H}_{SO} = \xi_{SO}\vec{L}\cdot\vec{S} = \xi_{SO}\left[-\frac{1}{\sqrt{2}}\left(V_{\kappa1}^{E_g} - V_{\kappa\breve{1}}^{E_g}\right) + +\frac{\imath}{\sqrt{2}}\left(V_{\mu1}^{E_g} + V_{\mu\breve{1}}^{E_g}\right) + V_{\nu0}^{A_{2g}}\right] \tag{6}$$

where the sign of $q$ is denoted by a breve symbol, i.e., $\breve{q} = -q$. Then we apply the Wigner-Eckart theorem to $V_{\lambda q}^{\Lambda}$, which implies working with reduced matrices $\langle\Gamma S||\mathbf{V}^{\Lambda}||\Gamma' S'\rangle$ [50]. In our case, these are 4×4 matrices defined in terms of the irreducible representations of the $D_{4h}$ point-group $\{\,^3E_g,\ ^3A_{2g},\ ^5A_{1g},\ ^5B_{1g}\}$, which are expressed as follows (see Appendix C for the details of this derivation):

$$\mathbf{V}^{A_{2g}} = \imath\begin{pmatrix} \sqrt{3} & 0 & 0 & 0\\ 0 & 0 & \sqrt{10} & 0\\ 0 & \sqrt{10} & 0 & 0\\ 0 & 0 & 0 & 0\end{pmatrix} \tag{7a}$$

$$\mathbf{V}^{E_g} = \imath\begin{pmatrix} 0 & \sqrt{3} & \sqrt{5} & -\sqrt{15}\\ \sqrt{3} & 0 & 0 & 0\\ \sqrt{5} & 0 & 0 & 0\\ -\sqrt{15} & 0 & 0 & 0\end{pmatrix} \tag{7b}$$

Once the reduced matrices are computed, the spin-orbit elements can be found using Clebsch-Gordan coefficients as follows:

$$\langle\,^3E_g\kappa M'|\vec{L}\cdot\vec{S}|\,^3E_g\mu M\rangle = -\frac{\imath}{2}\delta_{M'}^{M} \tag{8a}$$

$$\langle\,^3A_{2g}\nu M'|\vec{L}\cdot\vec{S}|\,^3E_g\kappa M\rangle = \frac{1}{2\sqrt{2}}\left[\delta_{M'}^{M+1} - \delta_{M'}^{M-1}\right] \tag{8b}$$

$$\langle\,^3A_{2g}\nu M'|\vec{L}\cdot\vec{S}|\,^3E_g\mu M\rangle = \frac{-\imath}{2\sqrt{2}}\left[\delta_{M'}^{M+1} + \delta_{M'}^{M-1}\right] \tag{8c}$$

$$\langle\,^3A_{2g}\nu M'|\vec{L}\cdot\vec{S}|\,^5A_{1g}uM\rangle = -\imath\sqrt{\frac{4-|M|}{3}}\delta_{M'}^{M} \tag{8d}$$

$$\langle\,^3E_g\kappa M'|\vec{L}\cdot\vec{S}|\,^5A_{1g}uM\rangle = \frac{\imath}{4}\sqrt{\frac{M^2+3|M|+2}{6}}\times \times\left[\delta_{M'}^{M+1} - \delta_{M'}^{M-1}\right] \tag{8e}$$

$$\langle\,^3E_g\mu M'|\vec{L}\cdot\vec{S}|\,^5A_{1g}uM\rangle = \frac{1}{4}\sqrt{\frac{M^2+3|M|+2}{6}}\times \times\left[\delta_{M'}^{M+1} + \delta_{M'}^{M-1}\right] \tag{8f}$$

$$\langle\,^3E_g\kappa M'|\vec{L}\cdot\vec{S}|\,^5B_{1g}vM\rangle = \frac{\imath}{4}\sqrt{\frac{M^2+3|M|+2}{2}}\times \times\left[\delta_{M'}^{M+1} - \delta_{M'}^{M-1}\right] \tag{8g}$$

$$\langle\,^3E_g\mu M'|\vec{L}\cdot\vec{S}|\,^5B_{1g}vM\rangle = -\frac{1}{4}\sqrt{\frac{M^2+3|M|+2}{2}}\times \times\left[\delta_{M'}^{M+1} + \delta_{M'}^{M-1}\right] \tag{8h}$$

With these relations, and taking the basis $\{|\,^3E_g\kappa1\rangle$, $|\,^3E_g\kappa0\rangle$, $|\,^3E_g\kappa\breve{1}\rangle$, $|\,^3E_g\mu1\rangle$, $|\,^3E_g\mu0\rangle$, $|\,^3E_g\mu\breve{1}\rangle$, $|\,^3A_{2g}\nu1\rangle$, $|\,^3A_{2g}\nu0\rangle$, $|\,^3A_{2g}\nu\breve{1}\rangle$, $|\,^5A_{1g}u2\rangle$, $|\,^5A_{1g}u1\rangle$, $|\,^5A_{1g}u0\rangle$, $|\,^5A_{1g}u\breve{1}\rangle$, $|\,^5A_{1g}u\breve{2}\rangle$, $|\,^5B_{1g}v2\rangle$, $|\,^5B_{1g}v1\rangle$, $|\,^5B_{1g}v0\rangle$, $|\,^5B_{1g}v\breve{1}\rangle$, $|\,^5B_{1g}v\breve{2}\rangle\}$, where again the breve symbol denotes the sign of spin quantum numbers, one can write the full 19x19-dimensional spin-orbit matrix as:

$$\mathcal{H}_{SO} = \frac{\xi_{SO}}{4\sqrt{3}}\times \times\left(\begin{array}{ccc:ccc|ccc|ccccc|ccccc}
0 & 0 & 0 & -\imath2\sqrt{3} & 0 & 0 & 0 & -\sqrt{6} & 0 & -\imath\sqrt{6} & 0 & \imath & 0 & 0 & -\imath3\sqrt{2} & 0 & \imath\sqrt{3} & 0 & 0\\
0 & 0 & 0 & 0 & 0 & 0 & \sqrt{6} & 0 & -\sqrt{6} & 0 & -\imath\sqrt{3} & 0 & \imath\sqrt{3} & 0 & 0 & -\imath3 & 0 & \imath3 & 0\\
0 & 0 & 0 & 0 & 0 & \imath2\sqrt{3} & 0 & \sqrt{6} & 0 & 0 & 0 & -\imath & 0 & \imath\sqrt{6} & 0 & 0 & -\imath\sqrt{3} & 0 & \imath3\sqrt{2}\\
\hdashline
\imath2\sqrt{3} & 0 & 0 & 0 & 0 & 0 & 0 & \imath\sqrt{6} & 0 & \sqrt{6} & 0 & 1 & 0 & 0 & -3\sqrt{2} & 0 & -\sqrt{3} & 0 & 0\\
0 & 0 & 0 & 0 & 0 & 0 & \imath\sqrt{6} & 0 & \imath\sqrt{6} & 0 & \sqrt{3} & 0 & \sqrt{3} & 0 & 0 & -3 & 0 & -3 & 0\\
0 & 0 & -\imath2\sqrt{3} & 0 & 0 & 0 & 0 & \imath\sqrt{6} & 0 & 0 & 0 & 1 & 0 & \sqrt{6} & 0 & 0 & -\sqrt{3} & 0 & -3\sqrt{2}\\
\hline
0 & \sqrt{6} & 0 & 0 & -\imath\sqrt{6} & 0 & & & & 0 & -\imath4\sqrt{3} & 0 & 0 & 0 & & & & & \\
-\sqrt{6} & 0 & \sqrt{6} & -\imath\sqrt{6} & 0 & -\imath\sqrt{6} & & 0 & & 0 & 0 & -\imath8 & 0 & 0 & & & 0 & & \\
0 & -\sqrt{6} & 0 & 0 & -\imath\sqrt{6} & 0 & & & & 0 & 0 & 0 & -\imath4\sqrt{3} & 0 & & & & & \\
\hline
\imath\sqrt{6} & 0 & 0 & \sqrt{6} & 0 & 0 & 0 & 0 & 0 & & & & & & & & & & \\
0 & \imath\sqrt{3} & 0 & 0 & \sqrt{3} & 0 & \imath4\sqrt{3} & 0 & 0 & & & & & & & & & & \\
-\imath & 0 & \imath & 1 & 0 & 1 & 0 & \imath8 & 0 & & & 0 & & & & & 0 & & \\
0 & -\imath\sqrt{3} & 0 & 0 & \sqrt{3} & 0 & 0 & 0 & \imath4\sqrt{3} & & & & & & & & & & \\
0 & 0 & -\imath\sqrt{6} & 0 & 0 & \sqrt{6} & 0 & 0 & 0 & & & & & & & & & & \\
\hline
\imath2\sqrt{3} & 0 & 0 & -3\sqrt{2} & 0 & 0 & & & & & & & & & & & & & \\
0 & \imath3 & 0 & 0 & -3 & 0 & & & & & & & & & & & & & \\
-\imath\sqrt{3} & 0 & \imath\sqrt{3} & -\sqrt{3} & 0 & -\sqrt{3} & & 0 & & & & 0 & & & & & 0 & & \\
0 & -\imath3 & 0 & 0 & -3 & 0 & & & & & & & & & & & & & \\
0 & 0 & -\imath2\sqrt{3} & 0 & 0 & -2\sqrt{3} & & & & & & & & & & & & &
\end{array}\right) \tag{9}$$

In this expression, solid lines separate the matrix elements corresponding to spectroscopic terms $\{\,^3E_g,\ ^3A_{2g},\ ^5A_{1g},\ ^5B_{1g}\}$ ordered from left to right columns. On the other hand, dotted lines separate the orbital angular momentum components ($\gamma = \kappa, \mu$) for the $^3E_g$ term. Finally the spin projections $M$ of the different elements are displayed in decreasing order from left to right.

We note that matrix Eq. (9) is represented for the quantization of $\vec{L}$ and $\vec{S}$ along the same axis. However, in general, the quantum spin axis can be oriented along arbitrary directions with respect to $\vec{L}$. Therefore, it is convenient to apply appropriate rotations $\mathcal{R}$ in the spin space to orient the spin quantization axis along arbitrary directions defined by $\hat{n}$ as follows:

$$S'_z = \mathcal{R} S_z \mathcal{R}^\dagger = \hat{n} \cdot \vec{S} \tag{10}$$

This rotation is characterized by an axis $\hat{t} = (\hat{z} \times \hat{n})/|\hat{z} \times \hat{n}|$ and a rotation angle $\theta = \arccos \hat{z} \cdot \hat{n}$.

$$\mathcal{R} = \mathrm{e}^{-\imath \theta \hat{t} \cdot \vec{S}} \tag{11}$$

If we define $\hat{n} = (\sin\theta \cos\phi, \sin\theta \sin\phi, cos\theta)$, we have $\hat{t} = (-\sin\phi, \cos\phi)$. Then, we obtain the following spin-rotation matrices for the cases $S = 1$ and $S = 2$:

$$\mathcal{R}^{(S=1)} = \frac{1}{2} \begin{pmatrix} 1+\cos\theta & \sqrt{2}\mathrm{e}^{\imath\phi}\sin\theta & \mathrm{e}^{2\imath\phi}(1-\cos\theta) \\ -\sqrt{2}\mathrm{e}^{-\imath\phi}\sin\theta & 2\cos\theta & \sqrt{2}\mathrm{e}^{\imath\phi}\sin\theta \\ \mathrm{e}^{-2\imath\phi}(1-\cos\theta) & -\sqrt{2}\mathrm{e}^{-\imath\phi}\sin\theta & 1+\cos\theta \end{pmatrix} \tag{12a}$$

$$\mathcal{R}^{(S=2)} = \frac{1}{8} \begin{pmatrix} 2(1+cos\theta)^2 & 4\mathrm{e}^{\imath\phi}\sin\theta(1+\cos\theta) & 2\sqrt{6}\mathrm{e}^{2\imath\phi}(\sin\theta)^2 & 4\mathrm{e}^{3\imath\phi}\sin\theta(1-\cos\theta) & 2\mathrm{e}^{4\imath\phi}(1-cos\theta)^2 \\ -4\mathrm{e}^{-\imath\phi}\sin\theta(1+\cos\theta) & [(1+4\cos\theta)^2-9]/2 & 4\sqrt{6}\mathrm{e}^{\imath\phi}\sin\theta\cos\theta & -\mathrm{e}^{2\imath\phi}[(1-4\cos\theta)^2-9]/2 & 4\mathrm{e}^{3\imath\phi}\sin\theta(1-\cos\theta) \\ 2\sqrt{6}\mathrm{e}^{-2\imath\phi}(\sin\theta)^2 & -4\sqrt{6}\mathrm{e}^{-\imath\phi}\sin\theta\cos\theta & 4[3(\cos\theta)^2-1] & 4\sqrt{6}\mathrm{e}^{\imath\phi}\sin\theta\cos\theta & 2\sqrt{6}\mathrm{e}^{-2\imath\phi}(\sin\theta)^2 \\ -4\mathrm{e}^{-3\imath\phi}\sin\theta(1-\cos\theta) & -\mathrm{e}^{-2\imath\phi}[(1-4\cos\theta)^2-9]/2 & -4\sqrt{6}\mathrm{e}^{-\imath\phi}\sin\theta\cos\theta & [(1+4\cos\theta)^2-9]/2 & 4\mathrm{e}^{\imath\phi}\sin\theta(1+\cos\theta) \\ 2\mathrm{e}^{-4\imath\phi}(1-cos\theta)^2 & -4\mathrm{e}^{-3\imath\phi}\sin\theta(1-\cos\theta) & 2\sqrt{6}\mathrm{e}^{-2\imath\phi}(\sin\theta)^2 & -4\mathrm{e}^{-\imath\phi}\sin\theta(1+\cos\theta) & 2(1+cos\theta)^2 \end{pmatrix} \tag{12b}$$

We use these matrices to compute the spin-orbit elements of Eq. (8) for arbitrary directions of the quantized spin axis.

## III. THEORY: INTERACTION WITH ELECTROMAGNETIC FIELDS

### A. Light-induced electron transfer between lattice sites

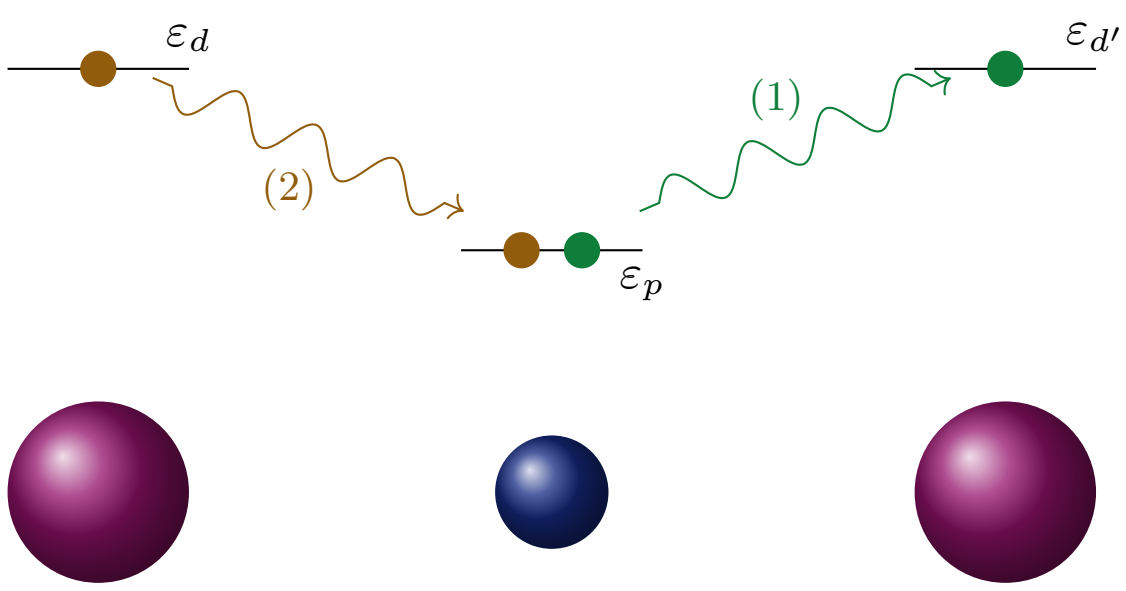

FIG. 3. Diagram of the electron transfer between neighbouring sites mediated by oxygen ions, induced by the interaction with an electromagnetic field. In the process, there is a first transition from oxygen p-orbital to neighbouring manganese d'-orbital. A second transition involves a transfer from a manganese d-orbital to an oxygen p-orbital.

So far we have considered on-site interactions of electronic orbitals with the crystal field, Jahn-Teller vibrational modes and atomic spin-orbit coupling. In the following, we describe their interaction with electromagnetic fields, which we assume induce electron transfer between neighboring sites in the perovskite lattice. The idea of light-induced electron transfer has been proposed, e.g., in some manganites, where optical energy excitations have been associated to polaronic transport due to intersite $e_g - e_g$ photoinduced transitions [25, 51, 52]. Since the cation separation in perovskites ($\approx 4$Å) is large for significant direct overlap [53–55], we consider the electron transfer dominated by hopping through $p$ orbitals of oxygen. As depicted in Fig. 3, we consider the transfer between neighbouring $d^4$ and $d^3$ ions, which can be described as:

$$d^4 p^6 d^3 \to d^4 p^5 d^4 \to d^3 p^6 d^4 \tag{13}$$

In order to describe this transfer, we define the following many-electronic wavefunctions for the two neighbouring $d$ ions and the oxygen ligands:

$$|\Psi\rangle = |\,^{2S+1}\Gamma\gamma M\rangle |\,^1S\rangle |\,^4A_{2g}N\rangle \tag{14a}$$

$$|\Phi_{p_w}\rangle = |\,^{2S+1}\Gamma\gamma M\rangle |\,^2Pw \pm \tfrac{1}{2}\rangle |\,^{2S+1}\Gamma'\gamma' M\rangle \tag{14b}$$

$$|\Psi'\rangle = |\,^4A_{2g}N\rangle |\,^1S\rangle |\,^{2S+1}\Gamma'\gamma' M\rangle \tag{14c}$$

Eq. (14a) corresponds to the initial configuration $d^4p^6d^3$, where the $d^4$ ion is described by some of the $^{2S+1}\Gamma$ representations discussed in Sec. I, whereas $^4A_{2g}$ is the ground state for the $d^3$ ion, according to the Tanabe-Sugano diagram [1]. On the other hand, the ligand orbitals, which have filled shells, are described by $^1S$ and Eq. (14c) describes the final state of the transfer, where the spin part of the wavefunction is unchanged since light cannot interact directly with spins. Finally, Eq. (14b) is the intermediate state where the two transition metals have 4 $d$-electrons and there is a vacancy in the ligand in a $p_w$ orbital ($w \in \{x, y, z\}$). This intermediate state requires an energy equivalent to the charge transfer energy, $\Delta_{CT} \approx 4$eV for $Mn^{3+}$ [44]. Therefore, in the presence of an electromagnetic field, the orbitals $p$ and $d$ are coupled, so that the states $|\Psi\rangle$ and $|\Psi'\rangle$ are perturbed by the intermediate states $|\Phi_{p_w}\rangle$ as follows:

$$|\tilde{\Psi}\rangle \approx |\Psi\rangle - \imath\frac{t_{pd}}{2\Delta_{CT}}\hat{\epsilon}\cdot\sum_w \langle\Phi_{p_w}|\vec{\nabla}|\Psi\rangle|\Phi_{p_w}\rangle \tag{15a}$$

$$|\tilde{\Psi}'\rangle \approx |\Psi'\rangle - \imath\frac{t_{pd}}{2\Delta_{CT}}\hat{\epsilon}\cdot\sum_w \langle\Phi_{p_w}|\vec{\nabla}|\Psi'\rangle|\Phi_{p_w}\rangle \tag{15b}$$

where $t_{pd}/2$ is the $p-d$ hopping amplitude induced by the electromagnetic field, which allows nonzero matrix elements between states $|\tilde{\Psi}\rangle$ and $|\tilde{\Psi}'\rangle$.

We treat the interaction with light to first order, so that the amplitude of the light-induced transfer requires the computation of electromagnetic matrix elements that involve two-center integrals including the vector potential $-\imath\vec{\nabla}$ (defined in the Coulomb gauge):

$$P^{\psi}_{q\hat{\epsilon}w} = \left(\frac{1}{\imath a}\int \psi(\vec{r})\vec{\nabla}\phi_{p_w}(\vec{r}\pm a\hat{e}_q)\mathrm{d}\vec{r}\right)\cdot\hat{\epsilon} \tag{16}$$

where $\hat{e}_q$ indicates the hopping direction in the lattice, $\hat{\epsilon}$ is the unit vector along the orientation of the vector potential, $a$ is the lattice parameter and $\psi$, $\phi_{p_w}$ describe monoelectronic orbitals in the transition metal and oxygen, respectively, that are involved in the photoinduced transfer. We note that although the spectroscopic terms are given as combinations of Slater determinants, the vector potential in Eq. (16) is a one-body operator that acts only on the monoelectronic orbital where the transferred electron resides (see Appendix D for a detailed discussion of how one-body operators act on the many-electron wavefunctions). The matrix elements shown in Eq. (16) are therefore expressed in terms of monoelectronic functions $\psi$ and $\phi_{p_w}$. This derives from the properties of the one-body potential, whereby matrix elements such as $-\imath\langle\Phi_{p_w}|\vec{\nabla}|\Psi\rangle$, where $|\Phi_{p_w}\rangle$, $|\Psi\rangle$ are many-electron functions described by Eq. (14), can be rewritten as $-\imath\langle\psi|\vec{\nabla}|\phi_{p_w}\rangle$, where $|\phi_{p_w}\rangle$, $|\psi\rangle$ describe monoelectronic orbitals.

On the other hand, while the expression in Eq. (16) corresponds to a transfer from a $p$ to a $d$ orbital, the $d$ to $p$ transition is described by its complex conjugate $(P^{\psi}_{q\hat{\epsilon}w})^*$. Interestingly, it can be shown that expressions like $\partial_{\hat{\epsilon}}\varphi_{p_w}$ (with $\hat{\epsilon}$ along $\hat{x}$, $\hat{y}$, or $\hat{z}$) appearing in Eq. (16) can be expressed as linear combinations of Slater-Koster coefficients (see Chapter 7 in Ref. [31] for a derivation). For instance, for the vector potential along $\hat{\epsilon}||\hat{x}$, we make use of the following coefficients:

$$(sd\sigma) \equiv \frac{1}{a}\int \psi_{z^2}(\vec{r})\bar{\psi}_s(\vec{r}\pm a\hat{e}_z)\mathrm{d}\vec{r} \tag{17}$$

$$(dd\sigma) \equiv \frac{1}{a}\int \psi_{z^2}(\vec{r})\bar{\psi}_{z^2}(\vec{r}\pm a\hat{e}_z)\mathrm{d}\vec{r} \tag{18}$$

$$(dd\pi) \equiv \frac{1}{a}\int \psi_{xy}(\vec{r})\bar{\psi}_{xy}(\vec{r}\pm a\hat{e}_x)\mathrm{d}\vec{r} \tag{19}$$

$$(dd\delta) \equiv \frac{1}{a}\int \psi_{xy}(\vec{r})\bar{\psi}_{xy}(\vec{r}\pm a\hat{e}_z)\mathrm{d}\vec{r} \tag{20}$$

where $\psi_{z^2}$, $\psi_{xy}$ are wavefunctions for the monoelectronic states $|z^2\rangle$ and $|xy\rangle$, and $\bar{\psi}_s$, $\bar{\psi}_{z^2}$ and $\bar{\psi}_{xy}$ are effective wavefunctions which have the same symmetries as $s$, $d_{z^2}$ and $d_{xy}$ orbitals (see Ref. [31]). Table I displays all nonzero matrix elements for the vector potential along the three directions in space in terms of the coefficients $(sd\sigma)$, $(dd\sigma)$, $(dd\pi)$ and $(dd\delta)$.

The hopping amplitudes $\alpha_q^{\psi_i\phi_j}$ between $\psi_i$ and $\phi_j$ orbitals located at neighboring sites ($i, j$ such that $\vec{r}_i - \vec{r}_j \parallel \hat{e}_q$) are calculated perturbatively, taking into account the $p-d$ hopping $t_{pd}$ and the charge transfer energy $\Delta_{CT}$ between $p$ and $d$ orbitals [44]:

$$\alpha^{\psi_i\phi_j}_{\hat{\epsilon}q} = \frac{t^2_{pd}}{\Delta_{CT}}\sum_w P^{\phi_j}_{q\hat{\epsilon}w}\left(P^{\psi_i}_{q\hat{\epsilon}w}\right)^* \tag{21}$$

Since the electromagnetic field cannot interact directly with spins, the matrix elements of the electromagnetic operator $\mathcal{W}_{\hat{\epsilon}}$ have the following form:

$$\langle i\Gamma\gamma SM|\mathcal{W}_{\hat{\epsilon}}|j\Gamma'\gamma'S'M'\rangle = \alpha^{\psi_i\phi_j}_{\hat{\epsilon}q}\delta^{S'}_{S}\delta^{M'}_{M} \tag{22}$$

where $i$, $j$ refer to neighboring locations in the lattice. In the next section we explain how the hopping amplitudes depend on the light polarization, which is described by the unit polarization vector $\hat{\epsilon}$ along an arbitrary direction.

### B. Cooperative Jahn-Teller effects

Although we address the dynamics of electron transfer from isolated Jahn-Teller ions, we incorporate cooperative effects, known to be relevant in solids [39, 44, 56–61]. The reason is that the dynamics of ions is much slower than the electronic transfer rates, so that we assume that cooperative effects restrict the possible Jahn-Teller deformations of the neighbouring sites where the transferred electron can jump into (see Fig. 4). As discussed in Sec. II A, we consider Jahn-Teller modes of the $d^4$ ion described by angles $\vartheta_n = 2n\pi/3 + \delta\vartheta, n = 0, 1, 2$ and $\delta\vartheta \ll 2\pi/3$. In consequence, there are three possible orientations for the transfer across the six oxygen anions surrounding the initial $d$-site, namely along $\pm\hat{x}$, $\pm\hat{y}$ or $\pm\hat{z}$. Then, cooperative effects are incorporated by imposing restrictions on the hopping from an initial $d^4$ ion with tetragonal distortion along $\hat{z}$ (Jahn-Teller mode with angle $\vartheta_0$) to neighboring sites along the three directions (see Fig. 4). For instance, when $\delta\vartheta>0$ there is a contraction along $y$ axis, forcing neighbours on the $xy$ plane to be distorted along the $y$ direction with $\delta\vartheta<0$. On the other hand, when a site distorted along $z$ has $\delta\vartheta<0$, there is a slight contraction along $x$, so their neighbours are distorted along $x$ with $\delta\vartheta>0$. Our model considers the orbital ordering with maximum entropy, which consists in having half of the octahedra distorted along a particular direction (chosen to be $z$) and the rest is equally distributed among elongations along $x$ and $y$ [62]. The resulting orbital ordering is depicted in Fig. 4.

| $\chi$ | $w$ | $q=x$ | $q=y$ | $q=z$ |
|---|---|---|---|---|
| | | $\hat{\epsilon}=\hat{x}$ | | |
| $u$ | $x$ | $-\frac{1}{2}sd\sigma-\frac{1}{2\sqrt{3}}dd\sigma$ | $-\frac{1}{2}sd\sigma+\frac{1}{4\sqrt{3}}dd\sigma+\frac{\sqrt{3}}{8}dd\delta$ | $sd\sigma-\frac{1}{2\sqrt{3}}dd\sigma$ |
| $v$ | $x$ | $\frac{\sqrt{3}}{2}sd\sigma+\frac{1}{2}dd\sigma$ | $-\frac{\sqrt{3}}{2}sd\sigma+\frac{1}{4}dd\sigma-\frac{1}{8}dd\delta$ | $-\frac{1}{4}dd\delta$ |
| $\eta$ | $z$ | $\frac{1}{2}dd\pi$ | $\frac{1}{2}dd\delta$ | $\frac{1}{2}dd\pi$ |
| $\tau$ | $y$ | $\frac{1}{2}dd\pi$ | $\frac{1}{2}dd\pi$ | $\frac{1}{2}dd\delta$ |
| | | $\hat{\epsilon}=\hat{y}$ | | |
| $u$ | $y$ | $-\frac{1}{2}sd\sigma+\frac{1}{4\sqrt{3}}dd\sigma+\frac{\sqrt{3}}{8}dd\delta$ | $-\frac{1}{2}sd\sigma-\frac{1}{2\sqrt{3}}dd\sigma$ | $sd\sigma-\frac{1}{2\sqrt{3}}dd\sigma$ |
| $v$ | $y$ | $\frac{\sqrt{3}}{2}sd\sigma-\frac{1}{4}dd\sigma+\frac{1}{8}dd\delta$ | $-\frac{\sqrt{3}}{2}sd\sigma-\frac{1}{2}dd\sigma$ | $\frac{1}{4}dd\delta$ |
| $\zeta$ | $z$ | $\frac{1}{2}dd\delta$ | $\frac{1}{2}dd\pi$ | $\frac{1}{2}dd\pi$ |
| $\tau$ | $x$ | $\frac{1}{2}dd\pi$ | $\frac{1}{2}dd\pi$ | $\frac{1}{2}dd\delta$ |
| | | $\hat{\epsilon}=\hat{z}$ | | |
| $u$ | $z$ | $-\frac{1}{2}sd\sigma+\frac{1}{4\sqrt{3}}dd\sigma-\frac{\sqrt{3}}{8}dd\delta$ | $-\frac{1}{2}sd\sigma+\frac{1}{4\sqrt{3}}dd\sigma-\frac{\sqrt{3}}{8}dd\delta$ | $sd\sigma+\frac{1}{\sqrt{3}}dd\sigma$ |
| $v$ | $z$ | $\frac{\sqrt{3}}{2}sd\sigma-\frac{1}{4}dd\sigma-\frac{1}{8}dd\delta$ | $\frac{-\sqrt{3}}{2}sd\sigma+\frac{1}{4}dd\sigma+\frac{1}{8}dd\delta$ | 0 |
| $\zeta$ | $y$ | $\frac{1}{2}dd\delta$ | $\frac{1}{2}dd\pi$ | $\frac{1}{2}dd\pi$ |
| $\eta$ | $x$ | $\frac{1}{2}dd\pi$ | $\frac{1}{2}dd\delta$ | $\frac{1}{2}dd\pi$ |

TABLE I. Nonzero matrix elements $\imath P^{\psi}_{q\hat{\epsilon}w}$ for the vector potential along the three directions in space $\hat{\epsilon}=\hat{x},\hat{y},\hat{z}$ in terms of Slater-Koster coefficients.

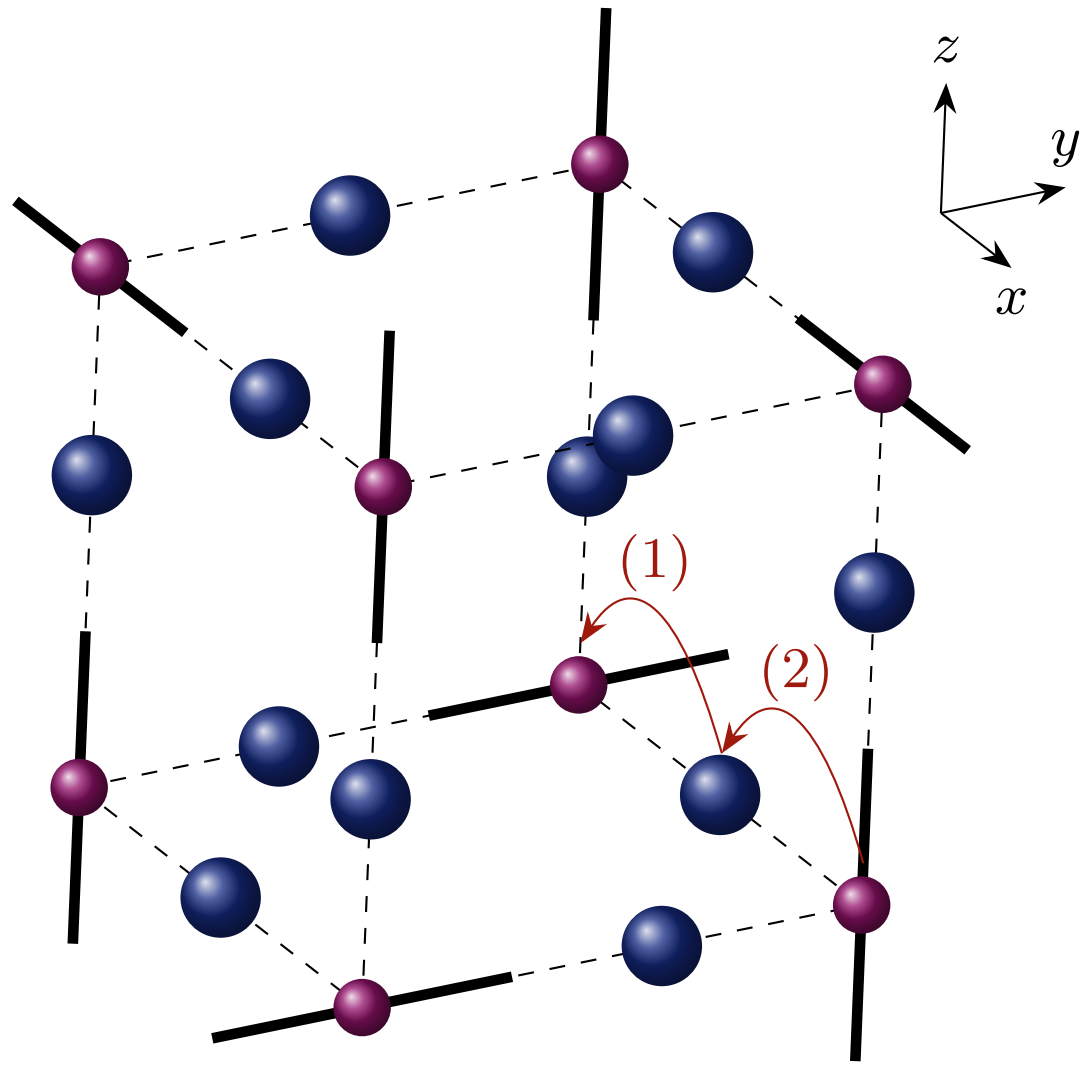

FIG. 4. Graphical representation of the cooperative distortions taking place during the light-induced transfer of electrons across the lattice. One of such transfers is illustrated by labels "1" and "2", where the electron jumps through an intervening oxygen. Each vertex represents a transition metal, around which the octahedron elongates along the solid lines. The electron is initially located at $d^4$ sites whose distortion is along $z$. Due to cooperative effects, the $d^3$ sites around the initial $d^4$ site are elongated along directions perpendicular to $z$.

### C. Hopping amplitudes for circularly polarized light

#### 1. *Left- and right- handed basis for circularly polarized light*

We describe the polarization of light by a complex vector $\hat{\epsilon} \in \mathbb{C}^3$ normalized to $\hat{\epsilon} \cdot \hat{\epsilon}^* = 1$:

$$\hat{\epsilon} = \frac{1}{\sqrt{E_{0x}^2 + E_{0y}^2 + E_{0z}^2}} \begin{pmatrix} E_{0x} \\ E_{0y} \mathrm{e}^{-\imath(\phi_y - \phi_x)} \\ E_{0z} \mathrm{e}^{-\imath(\phi_z - \phi_x)} \end{pmatrix} \tag{23}$$

Here $E_{0i}$ is the amplitude of the i-th component of the electric field and $\phi_i$ the correspondent phase, defined in this expression to keep $\epsilon_x$ real.

Since light is a transverse wave, we need only two complex vectors to define a basis for the polarization, both orthogonal to propagation direction $\hat{k}$. For circularly polarized light we use left-handed and right-handed polarizations, which, in the case of wave propagation along $\hat{k} = \hat{z}$ are defined as:

$$\hat{\varsigma}_L = \frac{1}{\sqrt{2}} \begin{pmatrix} 1 \\ \imath \\ 0 \end{pmatrix} \qquad \hat{\varsigma}_R = \frac{1}{\sqrt{2}} \begin{pmatrix} 1 \\ -\imath \\ 0 \end{pmatrix} \tag{24}$$

For arbitrary orientations of the propagation of light, we use the rotation matrix $R$ to find the new basis $\hat{\epsilon}_{L,R} = R\hat{\varsigma}_{L,R}$ for the polarization. This rotation is characterized by an axis $\hat{u} = (\hat{z} \times \hat{k})/\sin\alpha$ and an angle of rotation $\cos\varphi = \hat{z} \cdot \hat{k}$. Since every unit vector $\hat{k}$ can be described using polar and azimuth angles, $\alpha, \beta$, then $\hat{k} = (\sin\alpha\cos\beta, \sin\alpha\sin\beta, \cos\alpha)$, $\hat{u} = (-\sin\beta, \cos\beta, 0)$ and $\varphi = \alpha$. Then, the rotation matrix for arbitrary wave propagation can be defined as:

$$R = \begin{pmatrix} c_\alpha + s_\beta^2(1-c_\alpha) & -s_\beta c_\beta(1-c_\alpha) & s_\alpha c_\beta \\ -s_\beta c_\beta(1-c_\alpha) & c_\alpha + c_\beta^2(1-c_\alpha) & s_\alpha c_\beta \\ -s_\alpha c_\beta & -s_\alpha s_\beta & c_\alpha \end{pmatrix} \tag{25}$$

where a contracted notation for trigonometric functions is used, namely, $s_x = \sin x$ and $c_x = \cos x$. Finally the polarization vector for arbitrary wavevector orientation has the following expression:

$$\hat{\epsilon}_{L,R}(\hat{k}) = \frac{1}{\sqrt{2}} \begin{pmatrix} c_\alpha \mp \imath s_\beta(1-c_\alpha)\mathrm{e}^{\pm\imath\beta} \\ \pm\imath[c_\alpha + c_\beta(1-c_\alpha)\mathrm{e}^{\pm\imath\beta}] \\ -s_\alpha \mathrm{e}^{\pm\imath\beta} \end{pmatrix} \tag{26}$$

#### 2. *Electromagnetic response and time-reversal symmetry*

Left- and right-handed polarizations are related to each other by complex conjugation, $\hat{\epsilon}_L^* = \hat{\epsilon}_R$. With this relation we can deduce that $(P_{qLw}^{\psi})^* = -P_{qRw}^{\psi}$ (see Eq. (16)) if $\psi(\vec{r}) \in \mathbb{R}$, $\forall \vec{r} \in \mathbb{R}^3$, in other words: $\psi$ has real spatial symmetry, which is the case of the basis used here. Time reversal involves complex conjugation (since light acts only on the orbital angular momentum) and the interchange of the initial and final orbitals in the hopping, so that we have

$$(\alpha_{qL}^{\psi\phi})^* = \alpha_{qL}^{\phi\psi} = \alpha_{qR}^{\psi\phi} \tag{27}$$

which means that $\mathcal{K}\mathcal{W}_L\mathcal{K}^\dagger = \mathcal{W}_R$, being $\mathcal{K}$ the complex conjugation operator and $\mathcal{W}_{L,R}$ the electromagnetic operator for left- and right- handed light. We note that the Hamiltonian terms (see Eq. (2)) $\mathcal{H}_0$ and $\mathcal{H}_{JT}$ are expressed as real matrices, while $\mathcal{H}_{SO}$ is complex. Thus, in the absence of spin-orbit coupling ($\xi_{SO} = 0$), we have $\mathcal{K}\mathcal{H}_L\mathcal{K}^\dagger = \mathcal{H}_R$ and, as a consequence, the gyrotropic signal is zero. A nonzero gyrotropic response (a different response to electromagnetic waves of opposite handedness) arises only when $\xi_{SO} \neq 0$, which implies $\mathcal{K}\mathcal{H}_R\mathcal{K}^\dagger \neq \mathcal{H}_L$. This conclusion does not depend on the basis, since it holds even when the wavefunctions are not real, for instance, when they are expressed in spherical basis. Indeed, transforming from spherical to a real basis involves a unitary transformation $\mathcal{U}$, so that the relation between $\mathcal{W}_L$ and $\mathcal{W}_R$ is

$$\mathcal{U}^\dagger\mathcal{K}\mathcal{U}\mathcal{W}_L\mathcal{U}^\dagger\mathcal{K}\mathcal{U} = \tilde{\mathcal{K}}\mathcal{W}_L\tilde{\mathcal{K}}^\dagger = \mathcal{W}_R \tag{28}$$

where $\tilde{\mathcal{K}}$ is also an antiunitary transformation that keeps invariant $\mathcal{H}_0$ and $\mathcal{H}_{JT}$. This confirms that a change of basis does not break the time-reversal invariant relation between $\mathcal{H}_L$ and $\mathcal{H}_R$ when $\xi_{SO} = 0$.

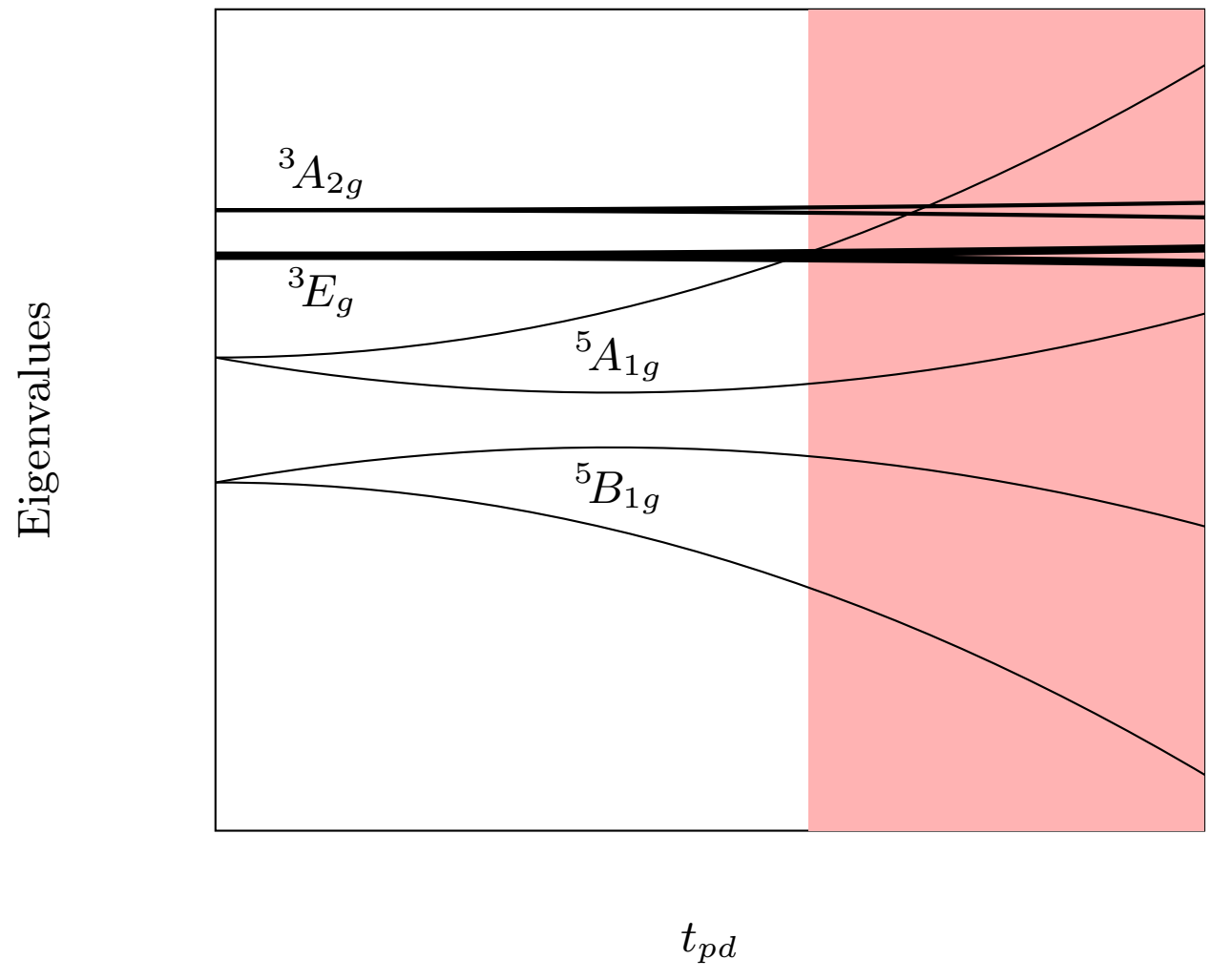

FIG. 5. Schematic depiction of the dependence on the $t_{pd}$ hopping of the eigenvalues of the wavefunctions corresponding to the irreducible representations $^5A_{1g}$, $^5B_{1g}$, $^3E_g$ and $^3A_{2g}$. The shadowed area corresponds to the parameter space where the spin-orbit mixing between $t_{2g}$ and $e_g$ states and, consequently, the gyrotropic response, are both strong.

#### 3. *Orbital-selective gyrotropic responses in broken time-reversal symmetry*

The origin of the gyrotropic responses can be traced back to the transfer induced by light between specific orbitals in the $d$ manifold. To shed light on this issue, it is convenient to express the polarization as $\hat{\epsilon}_L = (a, b, c)$ and $\hat{\epsilon}_R = (a^*, b^*, c^*)$. With this, the transfer amplitudes $t^{\psi\phi}_{q\hat{\epsilon}}$ in Eq. (27) can be expanded as products of $P^{\psi}_{q\hat{\epsilon}\hat{w}}$ integrals (see Eq. (16)) as

$$\begin{aligned}\alpha^{\psi\phi}_{qL} = -\sum_w \big[&|a|^2 P^{\psi}_{qxw}P^{\phi}_{qxw} + |b|^2 P^{\psi}_{qyw}P^{\phi}_{qyw} + \\ &+|c|^2 P^{\psi}_{qzw}P^{\phi}_{qzw} + a^*bP^{\psi}_{qxw}P^{\phi}_{qyw} + \\ &+a^*cP^{\psi}_{qxw}P^{\phi}_{qzw} + ab^*P^{\psi}_{qyw}P^{\phi}_{qxw} + \\ &+b^*cP^{\psi}_{qyw}P^{\phi}_{qzw} + ac^*P^{\psi}_{qzw}P^{\phi}_{qxw} + \\ &+bc^*P^{\psi}_{qzw}P^{\phi}_{qyw}\big]\end{aligned} \tag{29}$$

By inspection of Eq. (22) one realizes that any transfer involving hopping between $e_g$ and $t_{2g}$ orbitals at neighboring sites is forbidden, since $S = 2$ and $S' = 1$ (or viceversa). Then, by taking into account Eq. (29) and the nonzero transfer integrals $P^{\psi}_{q\hat{\epsilon}w}$ displayed in Table I, one can verify that all hopping amplitudes involving hopping between neighboring $e_g - e_g$ orbitals are real and are consequently time-reversal invariant. Therefore, the transfer between $e_g - e_g$ orbitals cannot give a gyrotropic response, at least to first order perturbation in the electromagnetic field. On the other hand, the light-induced transfer between neighboring $t_{2g} - t_{2g}$ orbitals has complex amplitude and breaks time-reversal symmetry, causing distinct electromagnetic responses for light of opposite handedness. As a consequence, both the presence of spin-orbit coupling and intersite $t_{2g} - t_{2g}$ transfer are key ingredients to have a gyrotropic signal.

In the light of the previous discussion, one expects a strong influence of the $t_{pd}$ hopping integral on the gyrotropic signal. Since the overlapping integrals $(sd\sigma)$ and $(dd\sigma)$ are significantly larger than $(dd\pi)$ and $(dd\delta)$, the energy of $t^3_{2g}e^1_g$ states (corresponding to $^5A_{1g}$ and $^5B_{1g}$ representations) is influenced much more strongly by $t_{pd}$ than $t^4_{2g}$ states with $^3E_g$ and $^3A_{2g}$ representations (see Fig. 5). It is then expected that as the value of $t_{pd}$ grows, the eigenvalues of $t^4_{2g}$ states will cross eventually the eigenvalues of $t^3_{2g}e^1_g$ states, producing a strong spin-orbit mixing and enhancing the gyrotropic signal (shadowed area in Fig. 5). In the present problem, we have verified numerically that this condition is fulfilled for values in the range $t^2_{pd}/\Delta_{CT} \sim 0.3\text{eV} - 1\text{eV}$.

## IV. PERTURBATION ANALYSIS OF THE GYROTROPIC RESPONSE

As discussed previously, a gyrotropic signal requires photoinduced transfer between adjacent $t_{2g} - t_{2g}$ orbitals. In consequence, the unperturbed ground state of the isolated Jahn Teller ion $^5B_{1g}$ $(t^3_{2g}e^1_g)$ has to be excited to a $t^4_{2g}$ configuration to activate this transfer channel. Here we derive a perturbation analysis in spin-orbit coupling and orthorhombic modes to understand the electronic transitions that contribute to the gyrotropic signal. First of all, we introduce the notation $|i\ ^{2S+1}\Gamma\gamma M\rangle$, which indicates the irreducible representation of the wavefunction at the i-th site in the lattice. The introduction of intersite hopping by interaction with the electromagnetic field breaks the degeneracy between the same state at different sites, originating 0-th order eigenstates denoted by:

$$|\alpha\ ^{2S+1}\Gamma\gamma M\rangle = \sum_i c^{\alpha}_i |i\ ^{2S+1}\Gamma\gamma M\rangle \tag{30}$$

To continue with the perturbed states we have to understand how the Hamiltonian acts on those 0-th order eigenstates. Recalling that the Hamiltonian has on-site $\mathcal{H}_S$ and inter-site electromagnetic $\mathcal{W}$ terms, we have:

$$\begin{aligned}&\langle\alpha'\ ^{2S'+1}\Gamma'\gamma' M'|\mathcal{H}_S|\alpha\ ^{2S+1}\Gamma\gamma M\rangle = \\ &\quad = \sum_i (c'^{\alpha'}_i)^* c^{\alpha}_i \langle i\ ^{2S'+1}\Gamma'\gamma' M'|\mathcal{H}_S|i\ ^{2S+1}\Gamma\gamma M\rangle\end{aligned} \tag{31a}$$

$$\begin{aligned}&\langle\alpha'\ ^{2S'+1}\Gamma'\gamma' M'|\mathcal{W}|\alpha\ ^{2S+1}\Gamma\gamma M\rangle = \\ &\quad = \sum_{i\neq j} (c'^{\alpha'}_i)^* c^{\alpha}_j \langle i\ ^{2S'+1}\Gamma'\gamma' M'|\mathcal{W}|j\ ^{2S+1}\Gamma\gamma M\rangle\end{aligned} \tag{31b}$$

A difficulty arises to compute such matrix elements due to cooperative Jahn-Teller effects. In particular, a given irreducible representation may contain different wavefunctions at adjacent sites in the lattice. For instance, the $^3A_{2g}$ term corresponds to a $|\zeta\eta\tau\bar{\tau}|$ Slater determinant if the distortion is along the $z$ direction, but if this distortion is along $y$ it corresponds to $|\zeta\eta\tau\bar{\eta}|$. Thus, the computation of matrix elements described by Eq. (31), necessary to determine the perturbed eigenstates, is challenging. We sort out this difficulty by approximating the matrix elements, for instance, $\langle\alpha'\ {}^3E_g t_{q'} M'|\mathcal{H}|\alpha\ {}^3E_g t_q M\rangle \sim \left(t_{pd}^2/\Delta_{CT}\right)(dd\pi)^2 + \xi_{SO} + F_T$. In this example, which can be generalized to arbitrary elements, every term is not determined exactly, but it gives a reasonable estimate of the contributions coming from intersite hopping, spin-orbit coupling and Jahn-Teller interactions.

Now we develop the perturbative analysis. Since we consider the dynamics of an electron initially located in a tetragonally elongated site, we are therefore interested in calculating the transition rates between $^5B_{1g}$ (corresponding to a $t_{2g}^3e_g^1$ configuration) and $^3E_g$ or $^3A_{2g}$ terms (both corresponding to a $t_{2g}^4$ configuration). By the effect of spin-orbit coupling and orthorhombic distortions, the non-perturbed wavefunctions $|\alpha\ {}^3A_{2g}\nu M\rangle$ and $|\alpha\ {}^3E_g t_q M\rangle$ become, respectively, $|\alpha\ {}^3A_{2g}\nu M\rangle^\bullet$ and $|\alpha\ {}^3E_g t_q M\rangle^\bullet$, where spherical harmonics are used to describe the orbital components $t_q$ of the wavefunctions (see Eq. (A10) for the definition of $t_q$). The matrix elements are then approximated to first order in spin-orbit coupling and orthorhombic modes as follows:

$$\langle\ {}^5B_{1g}v(\pm1-q)|\ {}^3E_g t_q(\pm1)\rangle^\bullet \sim \frac{\xi_{SO}}{\mathcal{E}_{EB}} \tag{32a}$$

$$\langle\ {}^5B_{1g}v(\pm1+q)|\ {}^3E_g t_q(\pm1)\rangle^\bullet \sim \frac{\delta\vartheta}{\mathcal{E}_{EB}}\left[F_T - \frac{\xi_{SO}}{\mathcal{E}_{EA}}\left(F_E - 2G_E \pm \frac{t_{pd}^2}{\Delta_{CT}}\frac{(sd\sigma)^2}{\delta\vartheta}\right)\right] \tag{32b}$$

$$\langle\ {}^5B_{1g}v(\pm1)|\ {}^3E_g\kappa 0\rangle^\bullet \sim \frac{\xi_{SO}}{\mathcal{E}_{EB}}\left[1 + \frac{\delta\vartheta}{\mathcal{E}_{EA}}\left(F_E - 2G_E \pm \frac{t_{pd}^2}{\Delta_{CT}}\frac{(sd\sigma)^2}{\delta\vartheta}\right)\right] \tag{32c}$$

$$\langle\ {}^5B_{1g}v(\pm1)|\ {}^3E_g\mu 0\rangle^\bullet \sim \frac{\xi_{SO}}{\mathcal{E}_{EB}}\left[1 - \frac{\delta\vartheta}{\mathcal{E}_{EA}}\left(F_E - 2G_E \pm \frac{t_{pd}^2}{\Delta_{CT}}\frac{(sd\sigma)^2}{\delta\vartheta}\right)\right] \tag{32d}$$

$$\langle\ {}^5B_{1g}vM|\ {}^3A_{2g}\nu M\rangle^\bullet \sim \frac{\xi_{SO}\delta\vartheta}{\mathcal{E}_{AB}\mathcal{E}_{AA}}\left(F_E - 2G_E \pm \frac{t_{pd}^2}{\Delta_{CT}}\frac{(sd\sigma)^2}{\delta\vartheta}\right) \tag{32e}$$

where $\mathcal{E}_{EB}$ and $\mathcal{E}_{EA}$ are, respectively, the energy gaps between $^3E_g$ and $^5B_{1g}$ and between $^3E_g$ and $^5A_{1g}$, while $\mathcal{E}_{AB}$ and $\mathcal{E}_{AA}$ are the analogous gaps corresponding to $^3A_{2g}$ instead of $^3E_g$. These energy gaps determine the degree of orbital mixing between $e_g$ and $t_{2g}$ states. The vibronic constants $(F_E, G_E, F_T)$, spin-orbit coupling $(\xi_{SO})$ and intersite hopping $(t_{pd})$ are also included in the expressions above. According to this perturbational analysis, the different transitions contributing to the gyrotropic signal are sketched in Fig. 6. We first note that spin-orbit corrections connect $^5B_{1g}$ with $^3E_g$, giving rise to the matrix element of Eq. (32a). On the other hand, Eq. (32b) stems from inter-site hoppings and orthorhombic corrections connecting $^5B_{1g}$ and $^5A_{1g}$ followed by spin-orbit mixing of $^5A_{1g}$ with $^3E_g$, while Eq. (32e) takes account of inter-site hoppings and orthorhombic corrections connecting $^5B_{1g}$ and $^5A_{1g}$ plus spin-orbit coupling between $^5A_{1g}$ and $^3A_{2g}$. Finally, Eq. (32c) and Eq. (32d) come from spin-orbit interactions within the $^3E_g$ subspace with $M = 0$, where the degeneracy of the wave-functions, which is preserved by spin-orbit coupling, lifts under the action of orthorhombic modes. An inspection of these expressions allows to extract the following conclusions:

- A relevant gyrotropic signal appears when the $e_g - t_{2g}$ spin-orbit mixing is large. According to Fig. 5, this happens when the gap $\mathcal{E}_{EA}$ between $^5A_{1g}$ and $^3E_g$ is reduced by the effect of light induced transfer through the $t_{pd}$ hopping integral. In this case, the strong reduction of $\mathcal{E}_{EA}$ entails an enhancement of contributions described by Eqs. Eq. (32b)-Eq. (32e).

- All amplitudes involving $^5B_{1g}$, $^3E_g$ and $^3A_{2g}$ described by Eq. (32) imply transitions between $t_{2g}^3e_g^1$ and $t_{2g}^4$ configurations, where one of the spins is inverted during the transition. The only exception is the transition described by Eq. (32e), which is a second-order correction in $\xi_{SO}\delta\vartheta$, i.e., it requires the simultaneous action of spin-orbit and orthorhombic interactions. Since $\delta\vartheta$ is small, and considering typical values for the vibronic constants $(F_T, F_E, G_E$. see section Sec. VI), the contribution of this term is negligible. Therefore, we conclude that the observation of a large gyrotropic signal is fundamentally contributed by transitions that in-volve a spin reversal.

- The perturbative influence of orthorhombic Jahn-Teller modes is described by the parameter $\delta\vartheta$. For small values of the $^5A_{1g}-$ $^3E_g$ gap, i.e., $\mathcal{E}_{EA} \lessapprox \xi_{SO}$, the predominant transition contributing to the gyrotropic signal is given by Eq. (32b). In this case, in addition to spin-orbit coupling, the hopping between neighboring $t_{2g} - t_{2g}$ states and orthorhombic modes enhance the gyrotropic signal. However, since the inter-site hopping is far larger than the energy of the orthorhombic distortions, the dependence of the gyrotropic response on $\delta\vartheta$ is very weak. On the other hand, for large enough values of the gap $\mathcal{E}_{EA} > \xi_{SO}$, the transition described by Eq. (32a) becomes predominant, but its amplitude is significantly smaller than for the case $\mathcal{E}_{EA} \lessapprox \xi_{SO}$. We can then conclude that the role of orthorhombic perturbations is minor, at least in the regime where $F_T\delta\vartheta \lessapprox \xi_{SO}$ and, therefore, the gyrotropic response is dominated by transitions between $^5A_{1g}$ and $^3E_g$, where the wavefunctions are perturbed by spin-orbit coupling.

We end this section by discussing the effects of the geometry on the gyrotropic signal, stemming from the relative orientations of light propagation and spin quantization, taking $\hat{z}$ as the orientation along the tetragonal distortion. An inspection of Eq. (29) reveals that for light propagating along $\hat{k} = \hat{z}$, namely, when light propagates along the distortion, the allowed gyrotropic hopping channel is mediated by $\zeta - \eta$ orbitals. In contrast, when light propagates perpendicular to the Jahn-Teller distortion the allowed gyrotropic hopping is $\eta - \tau$ for $\hat{k} = \hat{x}$, while for propagation along $\hat{k} = \hat{y}$ the gyrotropic hopping is mediated by $\tau - \zeta$ orbitals (see Eq. (A3c), Eq. (A3c) and Eq. (A3e) for a definition of the $t_{2g}$ orbitals $\zeta$, $\eta$ and $\tau$). As a result, the magnitude of the gyrotropic signal strongly depends on both spin axis and light propagation. The reason is as follows. In general, for a given couple of $t_{2g}$ orbitals in the hopping channel, the matrix elements of the angular momentum are nonzero only if the direction of the momentum component is contained in both spatial symmetries of the orbitals. For instance, for $\tau \sim xy$ and $\eta \sim xz$ orbitals, the only non-vanishing element is $\langle\tau|l_x|\eta\rangle \neq 0$. In addition, for a given pair of coupled orbitals in the hopping channel, it can be shown that the spin axis has to be oriented along the component of the nonzero matrix element to have a gyrotropic signal. For instance, for light propagating along the Jahn-Teller distortion, i.e., for $\hat{k} = \hat{z}$, the only gyrotropic channel is $\zeta - \eta$, Therefore, if the spin is quantized along $x$, then $\langle\zeta|l_x|\eta\rangle = 0$ and the gyrotropic signal is completely extinguished. Numerical calculations, discussed below, have been performed to study systematically the effect of geometry on the response to circularly polarized light.

In the following, we introduce the formalism to evaluate the gyrotropic response (section Sec. V), which we use to perform numerical calculations based on the exact diagonalization of the full Hamiltonian. In Sec. VI we analyze the influence of spin-orbit coupling and orthorhombic Jahn-Teller modes on the gyrotropic response, which confirms the general tendencies discussed in this section.

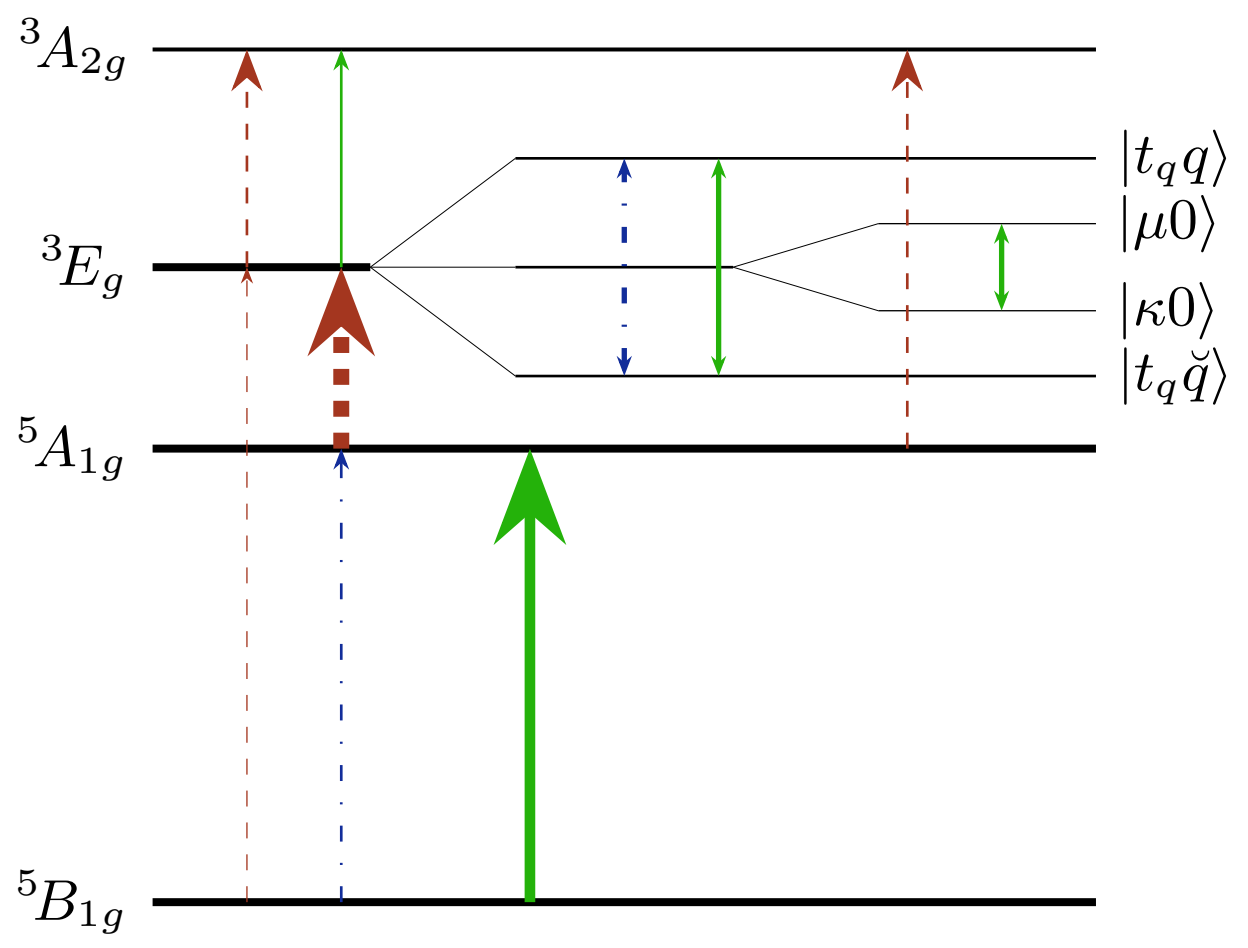

FIG. 6. Sketch of the transitions allowed by intersite hopping induced by light (green solid lines), spin-orbit coupling (brown dashed lines) and Jahn-Teller orthorhombic distortions (blue dashed-dotted lines). Thicker arrows indicate stronger interactions, corresponding to transitions between $^5B_{1g}$ and $^5A_{1g}$ mediated by intersite hopping and between $^5A_{1g}$ and $^3E_g$ mediated by spin-orbit coupling.

## V. RESPONSE TO CIRCULARLY POLARIZED ELECTROMAGNETIC WAVES

We consider the excitation of an electron located initially in a $d^4$ site distorted along $z$, see Fig. 4. In the presence of an electromagnetic field, this electron is transferred to any of the six nearest neighboring $d^3$ sites in the lattice. As explained in section Sec. III A, we assume that cooperative effects induce orbital ordering around the initial $d^4$ site, so that the site that receives the transferred electron can only deform along particular orientations, as shown in Fig. 4. In the calculations, the orbital ordering is extended periodically throughout the solid. To compute the dynamics, we suppose that the system is prepared in an ensemble

$$\varrho = \sum_{\psi} p_\psi |\psi\rangle\langle\psi| \tag{33}$$

where $\psi$ refers to state $^5B_{1g}$ in the central site, which has the lowest energy (see Fig. 1b). Here, $p_\psi$ is the relative weight assigned to each spin projection allowed by the irreducible representations of the corresponding many-electron wavefunctions. The values of $p_\psi$ are indeed obtained for each specific case after diagonalization of the full Hamiltonian. The ensemble in Eq. (33) evolves in time as

$$\varrho(t) = \mathrm{e}^{-\imath\mathcal{H}t}\varrho\,\mathrm{e}^{\imath\mathcal{H}t} \tag{34}$$

which allows us to compute the quantum propagator [63, 64]

$$G(t) = -\imath\Theta(t)\left\langle[\varrho(t), \varrho]\right\rangle \tag{35}$$

where the Heaviside function $\Theta(t)$ accounts for causality and $\langle\mathcal{O}\rangle$ is the thermal average of the operator $\mathcal{O}$.

$$\langle\mathcal{O}\rangle = \mathrm{tr}\left[\rho\mathcal{O}\right] \tag{36}$$

Here $\rho$ is the density matrix for a thermal bath

$$\rho = \frac{1}{Z}\sum_k \mathrm{e}^{-\beta\mathcal{H}}|k\rangle\langle k| = \frac{1}{Z}\sum_k \mathrm{e}^{-\beta\omega_k}|k\rangle\langle k| \tag{37}$$

where $Z$ is the partition function and the second equality holds if $\{|k\rangle\}$ is an eigenbasis of the Hamiltonian. We can then express the quantum propagator in the following way

$$\begin{aligned} G(t) = -\imath\Theta(t)\frac{1}{Z}\sum_{\substack{\psi\\k,m}} p_\psi^2\left(\mathrm{e}^{-\beta\omega_k} - \mathrm{e}^{-\beta\omega_m}\right)\times \\ \times\,\mathrm{e}^{\imath(\omega_m-\omega_k)t}|\langle k|\psi\rangle|^2|\langle m|\psi\rangle|^2 \end{aligned} \tag{38}$$

Here we have introduced the identity using the Hamiltonian eigenbasis labeled with $m$. Defining $\Omega_{km} = |\omega_k - \omega_m|$ and approximating $\mathrm{e}^{-\beta\omega} \approx 1$ for $\omega < \beta^{-1}$ and $\mathrm{e}^{-\beta\omega} \approx 0$ for $\omega > \beta^{-1}$ we can rewrite the propagator as follows preserving only the terms such that $\omega_k \gg \omega_m$:

$$\begin{aligned} G(t) = -\imath\Theta(t)\frac{1}{Z}\sum_{\psi} p_\psi^2 \sum_{k,m}\mathrm{e}^{-\imath\Omega_{km}t}\times \\ \times\,|\langle k|\psi\rangle|^2|\langle m|\psi\rangle|^2 + c.c. \end{aligned} \tag{39}$$

The first term in Eq. (39) corresponds to the retarded propagator, while the complex conjugate term is the advanced propagator. In order to compute the spectral response, we use the Heaviside function in the frequency domain

$$\Theta(t) = -\frac{1}{2\pi\imath}\lim_{\eta\to 0^+}\int_{-\infty}^{\infty}\mathrm{d}\omega\frac{\mathrm{e}^{-\imath\omega t}}{\omega + \imath\eta} \tag{40}$$

Using the expression Eq. (40) we can write the spectral representation of the propagator in frequency domain

$$G_r(\omega) = \frac{1}{Z}\sum_{\psi,m} p_\psi^2|\langle m|\psi\rangle|^2\sum_k\frac{|\langle k|\psi\rangle|^2}{\omega - \Omega_{km} + \imath\eta} \tag{41}$$

where $\Omega_{km}$ denotes the frequency eigenvalues of the full Hamiltonian. In the limit $\eta \to 0^+$, we have

$$\begin{aligned} \varsigma(\omega) = -\frac{1}{\pi}\Im[G_r(\omega)] = \frac{1}{Z}\sum_{\psi,m} p_\psi^2|\langle m|\psi\rangle|^2\times \\ \times\sum_k|\langle k|\psi\rangle|^2\delta(\omega - \Omega_{km}) \end{aligned} \tag{42}$$

which has the form of a density of states, which we denote as $\varsigma(\omega)$, while the parameter $\eta$ is related to the lifetime of the excited states. After some algebra, the spectral function can be rewritten as

$$\varsigma(\omega) = \frac{\eta}{Z\pi}\sum_{\psi,m} p_\psi^2|\langle m|\psi\rangle|^2\sum_k\frac{|\langle k|\psi\rangle|^2}{(\omega - \Omega_{km})^2 + \eta^2} \tag{43}$$

In the next section, we use this function to evaluate the gyrotropic response when time-reversal symmetry is broken.

## VI. NUMERICAL SIMULATIONS

### A. Calculation of the spectral functions for circularly polarized light

For the calculation of the spectral functions defined by Eq. (43) we have solved the full Hamiltonian (Eq. (1)) to compute the eigenvalues. The spectral functions have been obtained for left- ($\varsigma_L$) and right- ($\varsigma_R$) circularly polarized light, by calculating the hopping amplitudes as described in Sec. III C. From these functions, we have built non-gyrotropic ($\varsigma_{ng}$) and gyrotropic ($\varsigma_{gy}$) spectral functions in frequency space, which give account, respectively, of the dynamic responses that are insensitive and sensitive to the handedness of the polarization of light. These functions are defined as follows:

$$\varsigma_{ng}(\omega) = \frac{\varsigma_L(\omega) + \varsigma_R(\omega)}{2} \tag{44a}$$

$$\varsigma_{gy}(\omega) = \frac{\varsigma_L(\omega) - \varsigma_R(\omega)}{2} \tag{44b}$$

Finally, we define a function that integrates the gyrotropic signal over the analyzed spectral range (0eV – 3.5eV):

$$N_{gy} = \int_0^\infty \varsigma_{gy}(\omega)\mathrm{d}\omega \tag{45}$$

The numerical calculations were carried out by setting the vibronic constants to $F_E = 450$meV, $F_T = 130$meV and $G_E = 20$meV. These values are in agreement with the Jahn-Teller splitting observed for $e_g$ and $t_{2g}$ electrons in 3d elements [44]. On the other hand, the charge transfer gap has been set to $\Delta_{CT} = 4$eV [44], the damping factor to $\eta = 180$meV (see Eq. (43)) and the $p-d$ hopping to $t_{pd} = 1.2$eV (Eq. (21)). Finally, the Slater-Koster coefficients were set to $(sd\sigma) = 1$, $(dd\sigma) = 0.82$, $(dd\pi) = 0.29$ and $(dd\delta) = 0.07$.

We studied different geometries by varying the relative orientation of the light propagation, Jahn-Teller distortions and spin quantization. By way of illustration, the spectral functions $\varsigma_{ng}$ and $\varsigma_{gy}$ displayed in Fig. 7 were calculated for three different geometries, which are sketched in the top panels of each column. The spectral functions were computed for four different values of the energy gap $\Delta$ (as indicated in the panels of Fig. 7), while the spin-orbit coupling was set to $\xi_{SO} = 20$meV. The gap $\Delta$ is defined as the energy difference between the unperturbed ${}^5A_{1g}$ and ${}^3E_g$ terms (see Fig. 1b), which gives an estimation of the degree of $t_{2g} - e_g$ mixing before the introduction of the electromagnetic field. All parameters, including $\Delta$, were chosen to work in a region of the Tanabe-Sugano diagram appropriate for manganese ions, for which the crystal field is $10D_q \approx 2eV$ and the Racah parameter is $B \approx 0.11 - 0.13eV$ [26, 65, 66]. On the other hand, in Fig. 8, the integrated gyrotropic signal described by $N_{gy}$ (Eq. (45)) is mapped as a function of spin-orbit coupling $\xi_{SO}$ and orthorhombic perturbations $\delta\theta$ for each value of $\Delta$. Panels in Fig. 8 are organized in the same way as in Fig. 7, i.e., each column corresponds to each of the geometries sketched on the top.

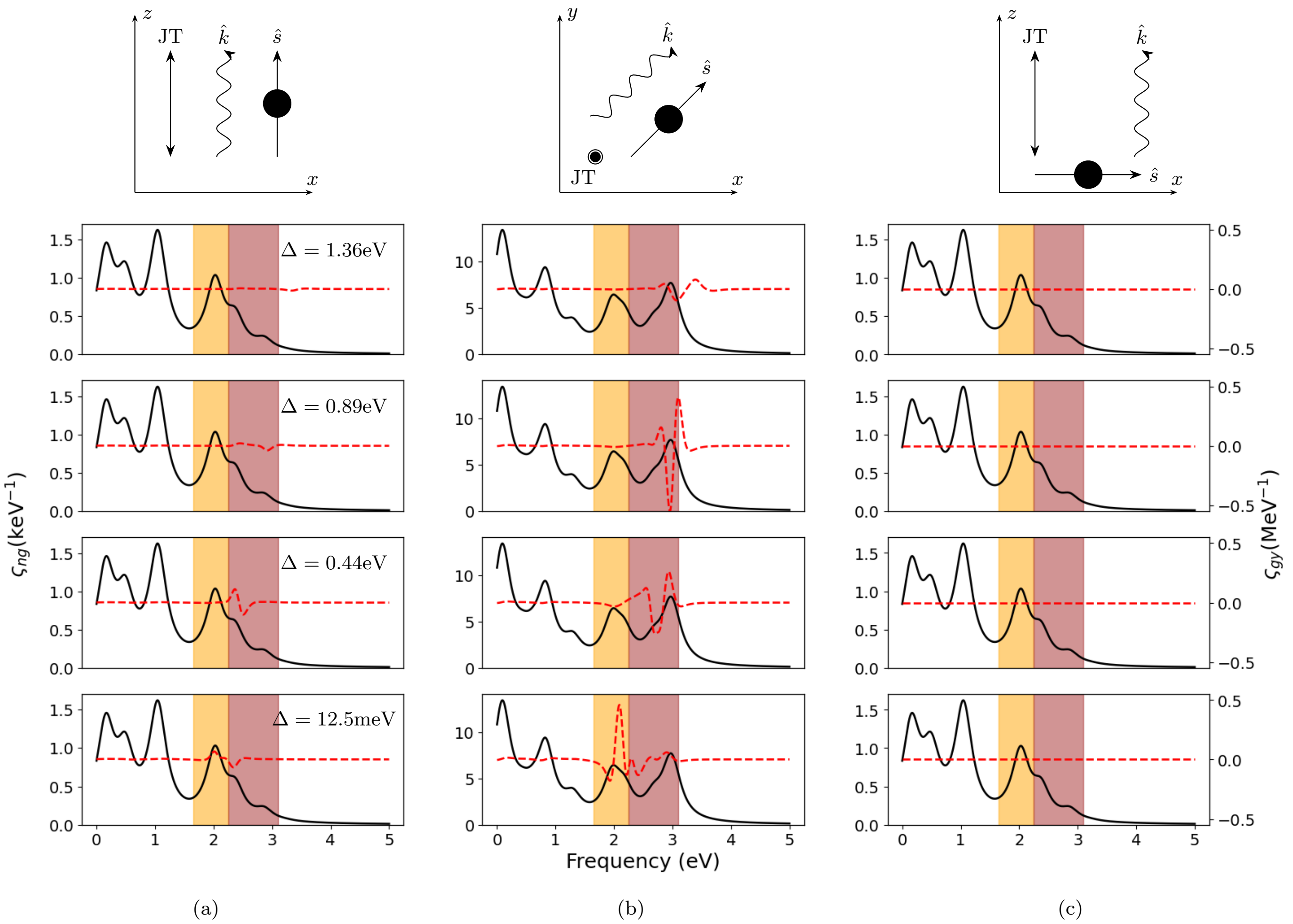

FIG. 7. Nongyrotropic ($\varsigma_{ng}$) (black solid lines) and gyrotropic ($\varsigma_{gy}$) (red dashed lines) spectral functions calculated for different energy gaps $\Delta$ defined between ${}^5A_{1g}$ and ${}^3E_g$ terms (see Fig. 1b). The spectra computed for $\Delta = 1.36\text{eV}, 0.89\text{eV}, 0.44\text{eV}, 12.5\text{meV}$ are displayed in descending order for each column (labelled a), b) and c)). The different functions have been computed in different geometric configurations, as sketched on the top of each column. The visible part of the spectrum has been shadowed and divided in two parts, below and above 550nm. This division enables an easier comparison with the experimental magneto-optical spectra reported in Ref. [25]. Those experiments show two absorption peaks centered, respectively, at wavelengths $<$ 550nm and $>$ 550nm, where only the latter gives rise to a gyrotropic response. The spectra computed in column b) for a gap $\Delta = 1.19$eV are in agreement with the experimental spectra reported in [25].

### B. Discussion of gyrotropic and nongyrotropic responses

We first discuss the nongyrotropic spectra described by functions $\varsigma_{ng}$ shown in Fig. 7. First of all, we observe that the structure of resonances remains virtually unchanged, as long as the geometry is fixed, regardless of the values of the other parameters. In addition, a comparison between the $\varsigma_{ng}$ spectra displayed in Fig. 7 (a) and (c) shows that the nongyrotropic response does not depend on the direction of the spin quantization $\hat{s}$, provided that the relative orientations of light propaga-tion and distortions remain the same. This is an indication that the nongyrotropic spectra are contributed essentially by transitions between $^5B_{1g}$ and $^5A_{1g}$ states. This observation is supported by the fact that the corresponding spin-orbit elements are zero for these states (see Eq. (9)), which explains why the $\varsigma_{ng}$ spectra remain unchanged as the axis of spin quantization changes. Therefore, the structure of resonances observed in $\varsigma_{ng}$ arises basically from the interactions between $^5B_{1g}$ and $^5A_{1g}$ states, mostly through intersite hopping induced by light (Fig. 5). More specifically, such transitions connect $^5B_{1g}$ and $^5A_{1g}$ bonding/antibonding states emerging from light-induced hybrization, which explains the structure of the peaks in the spectra (Fig. 7). The resonances located at lower energy, below the visible range (shadowed areas in Fig. 7) are mostly contributed by transitions between hybridized $^5B_{1g}$ orbitals, while the resonances located in the visible region correspond to transitions involving $^5A_{1g}$ states.

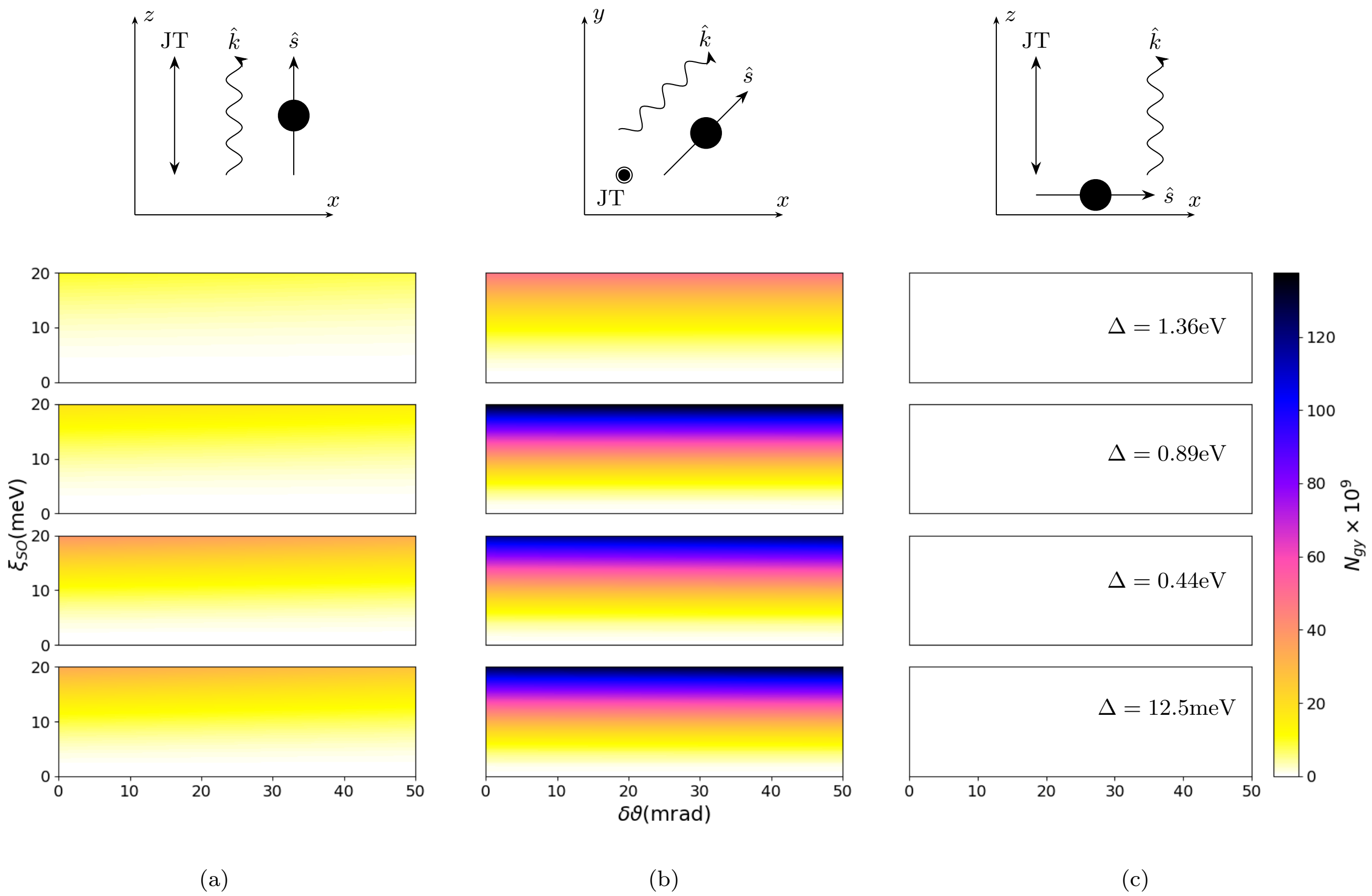

FIG. 8. Maps of the integrated spectral function $N_{gy}$ defined by Eq. (45), as a function of the spin-orbit coupling ($\xi_{SO}$) and orthorhombic perturbations ($\delta\vartheta$). The maps were computed for the different geometries sketched in the top of each column, labelled as a), b) and c). For each column, the integrated spectral functions $N_{gy}$ were calculated for different values of the energy gap $\Delta$.

We turn now our attention to the gyrotropic $\varsigma_{gy}$ spectra. As observed in Fig. 7, their structure is much simpler, with a main resonance located in the visible or near-infrared, depending on the geometry and value of $\Delta$. The contributions to this resonance come mainly from transitions between $^3E_g$ and $^3A_{2g}$ states that are perturbed by spin-orbit coupling (see section Sec. IV). The spectral weight of these transitions is too small to be observed in the nongyrotropic spectra. Nonetheless, their effect on the gyrotropic response is crucial, via the orbital mixing induced by spin orbit coupling. One way to evaluate this mixing is by varying the energy gap $\Delta$ defined above. In particular, we observe in Fig. 7 that the gyrotropic signal is the smallest for the largest value of $\Delta$, as expected from the smaller orbital mixing in this case.

Let us now discuss the effect of orbital hybridization induced by the coupling to the electromagnetic field. As discussed in Sec. IV, the coupling to light induces $p-d$ hybridization, which causes an evolution of the eigenstates as a function of the overlapping between oxygen and transition metal states, as sketched in Fig. 5. In our numerical calculations we observe that for values $t_{pd} \approx 1.2$eV and $\Delta \lessapprox 0.89$eV there is a strong orbital mixing. Indeed, for this choice of values, the calculated spectra are in agreement with the experimental spectra reported in Ref. [25]. In particular, in the calculated spectra we observe two nongyrotropic resonances in the red and blue parts of the visible range, respectively, while a main gyrotropic resonance is seen in the blue region, in agreement with the experiments [25].

Next we discuss the effects of the geometry on the gyrotropic response. The data shown in Fig. 7 and Fig. 8 reveals the strong dependence of the spectral functions on geometric factors. The effect is particularly evident for spectra shown in panels (c) of both Figures, which show that the gyrotropic signal is completely extinguished for this particular geometry. As discussed in Sec. IV, the reason for this extinction is that the matrix elements of the angular momentum that connect the $t_{2g}$ orbitals in adjacent sites are null in this case, because the direction of the momentum component is not contained in the space symmetries of the $t_{2g}$ orbitals involved in the transfer. On the other hand, the gyrotropic signal for the geometry sketched in panels (b) of both Figures is significantly larger than for the spectra and maps shown in panels (a). The reason is that for the geometry of spectra and mappings shown in Fig. 7(a) and Fig. 8(a), the photoinduced transfer that contributes to the gyrotropic response happens only between adjacent $\eta - \tau$ orbitals. In contrast, for the geometry studied in Fig. 7(b) and Fig. 8(b) both $\zeta - \eta$ and $\zeta - \tau$ hopping channels are allowed, increasing the gyrotropic signal.

Finally, the data shown in the maps of Fig. 8 reveals the dependence of the gyrotropic response on spin-orbit coupling. Indeed, when the latter tends to zero, the gyrotropic signal becomes vanishingly small, while it becomes progressively more intense as the spin-orbit coupling increases. On the other hand, we see that orthorhombic perturbations barely modify the gyrotropic spectra, as expected from the fact that these perturbations are much smaller than the intersite hopping induced by coupling to the electromagnetic field. Both observations are in agreement with the conclusions drawn from the perturbative analysis discussed in Sec. IV.

## VII. CONCLUSIONS AND PERSPECTIVES

We used a group theoretical approach to study the interaction of transition metals with electromagnetic fields. For that purpose, we described the relevant electronic states by irreducible representations of pertinent point-group symmetries, which were constructed from many-electron wavefunctions based on Slater determinants. The energetics of the problem was established to comply with Tanabe-Sugano diagrams corresponding to the particular ion under study [1]. Starting from an initial $O_h$ symmetry, we analyzed the effect of symmetry reduction due to Jahn-Teller interactions and spin-orbit coupling. The interaction with electromagnetic fields was assumed to produce photoinduced transfer of electrons across the lattice. The model predicts an electronic response that depends on the handedness of circular polarization of light. Key ingredients to this gyrotropic response are spin-orbit coupling and intraatomic $e_g - t_{2g}$ orbital mixing. Remarkably, the gyrotropic resonances are related to photoexcitations where one of the spins is inverted, enabling the use of electromagnetic fields to manipulate spins. We also analyzed the dependence of the gyrotropic response on the relative orientation of Jahn-Teller distortions, light propagation and spin quantization. In particular, we found specific conditions for which the gyrotropic response is largely reduced or even extinguished. We focused our analysis on $3d$ ions, with the aim of understanding our recent observation of a gyrotropic response associated with Jahn-Teller polarons in $La_{2/3}Ca_{1/3}MnO_3$ [25]. Using standard values for Jahn-Teller constants, spin-orbit coupling and charge transfer our model replicates a gyrotropic response in the blue region of the visible range, in agreement with the experiments [25].

Beyond this particular case, a similar approach may be generalized to study the interaction with electromagnetic fields of transition metals with arbitrary point symmetries and spin-orbit couplings. One perspective is the entanglement of spin and orbital degrees of freedom using light at optical wavelengths. One could think, for instance, of studying quantum tunneling of $E \otimes e$ or $T \otimes e$ Jahn-Teller vibronic states [37]. In particular, the interaction with light may drive photoexcited states, whose ground state is formed by coherent superpositions of those vibronic states through quantum tunneling, which may be detected with polarized light. Such excitations could form the basis for quantum states of interest for quantum technologies [22, 67]. Another prospect may be the study of 4d-5d transition metals hosting (quantum) spin liquids [18–21]. In this case, magnetic interactions would compete with the coupling to the electromagnetic field, which could lead to a rich diagram of quantum phases as a function of the wavelength of the electromagnetic radiation. To solve this problem, a group-theoretical approach would require working in appropriate regions of Tanabe-Sugano diagrams [1, 49]. For instance, for heavy metal $d^4$ ions the ground state is $^3T_{1g}$ instead of $^5E_g$, due to the larger crystal field that leads to the condition $(Dq/B)_c > 2.7$, see Figure 1a. Additionally, since in this case Jahn-Teller interactions in the $t_{2g}$ manifold are typically smaller than spin-orbit coupling [20, 37, 44], the group-theoretical analysis should consider lowering the point symmetry by spin-orbit interactions in the first place.

We note that our model Hamiltonian considers electrons that are subject to Jahn-Teller instabilities. In general, these may coexist with electrons in delocalized bands. This is indeed the case of many oxides, including $La_{2/3}Ca_{1/3}MnO_3$, where both Jahn-Teller polarons and delocalized electrons participate in transport [68–70]. We stress that our model captures the essential physics of electrons that are affected by Jahn-Teller interactions, neglecting contributions from delocalized electrons. As we demonstrate here, this is enough to describe the specific contribution of Jahn-Teller vibronic states to the gyrotropic response, which is experimentally distinguish-able from the conventional response arising from delocalized bands [25]. On the other hand, although the analysis based on spectral functions gives a fundamental understanding of light-matter interactions in these solids, further developments can address linear response theory to obtain responses like optical conductivity and permittivity that can be matched with experiments.

We also note that our assumption of photoinduced electron transfer implies an enhanced conductivity at resonant frequencies, which, eventually could be tested experimentally by measuring electronic transport under illumination at relevant wavelengths. These experiments could be done in $La_{2/3}Ca_{1/3}MnO_3$, but other candidates would also comprise materials like $(Pr_xLa_{1-x})_{2/3}Ca_{1/3}MnO_3$ [71] or magnetite [72], where optical signatures of Jahn-Teller polarons have been observed [73]. Generally, materials prone to Jahn-Teller instabilities, including colossal magnetoresistance manganites [68–70], could be worth exploring in search for gyrotropic responses arising from spin-orbital mixing. On the theoretical side, other models can extend the analysis to the optical responses of clusters of Jahn-Teller ions rather than isolated ions. Alternative models may also explore these responses in the absence of phototransfer, e.g., the photoexcitation of Jahn-Teller states in molecules [37, 74], in which the group-theoretical approach should be applied at the level of molecular orbitals. Finally, in the present model, the electromagnetic radiation and the lattice modes are treated as classical fields. Further extensions would require a full quantum approach to describe these fields, especially relevant for the application of the aforementioned ideas to concepts like cavity quantum electrodynamics [67, 75].

## ACKNOWLEDGMENTS

We acknowledge financial support from PID2020-118479RBI00 and Severo Ochoa FUNFUTURE (CEX2019-000917-S) projects of the Spanish Ministry of Science and Innovation (Grant No. MCIN/AEI/10.13039/501100011033).

## Appendix A: Wavefunctions of many-electron states

### 1. Notation

We use group theory to construct many-electron wavefunctions. We work in a product basis between orbital and spin momenta, so that wavefunctions are defined by kets like:

$$|(G)\ ^{2S+1}\Gamma(\gamma)M\rangle \tag{A1}$$

Here $S$ is the spin magnitude, so that $2S+1$ is the spin degeneracy, and $\Gamma$ is the irreducible representation in orbital space of group $G$, expressed in the basis $\gamma$ (which is omitted for unidimensional representations). Finally $M$ is referred to the spin quantum number, $M = -S, ..., S$. In the description of spin-orbit coupling, wavefunctions are denoted by $|\Gamma\gamma\rangle$ since orbital and spin angular momenta are coupled and double group representations are best suited to take into account the electron spin.

The wavefunctions describe multi-electronic states, so they are linear combinations of Slater determinants:

$$|\psi_1 \cdots \psi_N| = \frac{1}{\sqrt{N!}} \begin{vmatrix} \psi_1(1) & \cdots & \psi_N(1) \\ \vdots & \ddots & \vdots \\ \psi_1(N) & \cdots & \psi_N(N) \end{vmatrix} \tag{A2}$$

Here $\psi_i = \phi_i \chi_i$ is the i-th mono-electronic wavefunction, where the orbital angular momentum part is described by $\phi_i = \zeta, \eta, \tau, u, v$ (labels are defined in EqEq. (A3)) and $\chi_i = \alpha, \beta$ is the spinor part. Being $\hat{n}$ the quantization axis for the spin, then $\hat{n} \cdot \hat{s}\alpha = +1/2$ and $\hat{n} \cdot \hat{s}\beta = -1/2$. For brevity, the notation inside the Slater determinants is written as $\phi\alpha \to \phi$ and $\phi\beta \to \bar{\phi}$, which is taken from Ref. [1]. Finally, the sign in bras and kets for wavefunctions is denoted by a breve symbol, i.e., $\breve{M} = -M$.

### 2. Construction of the wavefunctions

#### a. $O_h$ point symmetry

In the case of one electron in a $d$ shell in a cubic crystal field the ten-fold degeneracy of the free ion is broken into a $e_g$ shell (with four-fold degeneracy) and a $t_{2g}$ shell (six-fold degeneracy). The basis angular functions of these two shells can be expressed as linear combinations of spherical harmonics $Y_l^m$ to get real spatial symmetries:

$$u = Y_2^0 \sim \frac{1}{2}(3z^2 - r^2) \tag{A3a}$$

$$v = \frac{1}{\sqrt{2}}\left[Y_2^{+2} + Y_2^{-2}\right] \sim \frac{\sqrt{3}}{2}(x^2 - y^2) \tag{A3b}$$

$$\zeta = \frac{\imath}{\sqrt{2}}\left[Y_2^{+1} + Y_2^{-1}\right] \sim \sqrt{3}yz \tag{A3c}$$

$$\eta = -\frac{1}{\sqrt{2}}\left[Y_2^{+1} - Y_2^{-1}\right] \sim \sqrt{3}xz \tag{A3d}$$

$$\tau = -\frac{\imath}{\sqrt{2}}\left[Y_2^{+2} - Y_2^{-2}\right] \sim \sqrt{3}xy \tag{A3e}$$

$u$ and $v$ are the basis for the $e_g$ shell and $\zeta$, $\eta$ and $\tau$ are the basis for the $t_{2g}$ shell.

When there is more than one electron in the $d$ shell we construct many-electron wavefunctions using Slater determinants. Since the spin-orbit interaction is small compared to exchange interactions, the many-electron wavefunctions are built by coupling separately the orbital and spin momenta, following the Russell-Saunders coupling scheme and using Clebsh-Gordan coefficients. For the orbital part, the coefficients are adapted for the $O_h$ point-group symmetry. The many-electron wavefunctions are expressed as linear combinations of Slater determinants, which take all possible permutations of electrons sitting on the different monoelectronic orbitals and having all possible spin orientations [1].

In the $d^3$ configuration the ground state for all values of the Racah parameter $\Delta/B$ in the Sugano-Tanabe diagrams is ${}^4A_{2g}$. It can be shown that the wavefunction that describes this four-fold degenerated term is built from Slater determinants with the three monoelectronic orbitals of $t_{2g}$ as follows [1]:

$$| \, {}^4A_{2g} \tfrac{3}{2} \rangle = -|\zeta\eta\tau| \tag{A4a}$$

$$| \, {}^4A_{2g} \tfrac{1}{2} \rangle = -\frac{1}{\sqrt{3}} \left[ |\zeta\eta\bar{\tau}| + |\zeta\bar{\eta}\tau| + |\bar{\zeta}\eta\tau| \right] \tag{A4b}$$

$$| \, {}^4A_{2g} \breve{\tfrac{1}{2}} \rangle = -\frac{1}{\sqrt{3}} \left[ |\bar{\zeta}\eta\bar{\tau}| + |\zeta\bar{\eta}\bar{\tau}| + |\bar{\zeta}\bar{\eta}\tau| \right] \tag{A4c}$$

$$| \, {}^4A_{2g} \breve{\tfrac{3}{2}} \rangle = -|\bar{\zeta}\bar{\eta}\bar{\tau}| \tag{A4d}$$

When a fourth electron is added to the $t_{2g}$ shell, the many-electron wavefunction transforms as a $T_{1g}$ representation with total spin $S = 1$ and is built from nine degenerate Slater determinants as follows:

$$| \, {}^3T_{1g}\kappa 1 \rangle = |\zeta\eta\tau\bar{\zeta}| \tag{A5a}$$

$$| \, {}^3T_{1g}\kappa 0 \rangle = \frac{1}{\sqrt{2}} \left[ |\zeta\bar{\eta}\tau\bar{\zeta}| + |\zeta\eta\bar{\tau}\bar{\zeta}| \right] \tag{A5b}$$

$$| \, {}^3T_{1g}\kappa \breve{1} \rangle = |\zeta\bar{\eta}\bar{\tau}\bar{\zeta}| \tag{A5c}$$

The orbital basis functions for $T_{1g}$ are $\kappa, \mu, \nu$ which have the same relation under rotations as the Cartesian coordinates $x, y, z$ but being even under parity. Eq. (A5) give us three determinants. The remaining six are obtained for the $\mu$ and $\nu$ basis of the ${}^3T_{1g}$ representation. The latter are obtained from $\kappa$ by rotating in the orbital space by $\pm 2\pi/3$ around the [111] axis.

On the other hand, if the fourth electron is in the $e_g$ shell the many-electron wavefunction corresponds to an $E_g$ representation. In this case, the spin number is $S = 2$ and the many-electron wavefunctions can be expressed as ten linear combinations of determinants as follows:

$$| \, {}^5E_g\gamma 2 \rangle = \pm|\zeta\eta\tau\gamma'| \tag{A6a}$$

$$\begin{aligned} | \, {}^5E_g\gamma 1 \rangle = \pm\frac{1}{2} \big[ & |\zeta\eta\tau\bar{\gamma}'| + |\bar{\zeta}\eta\tau\gamma'| + |\zeta\bar{\eta}\tau\gamma'| + \\ & + |\zeta\eta\bar{\tau}\gamma'| \big] \end{aligned} \tag{A6b}$$

$$\begin{aligned} | \, {}^5E_g\gamma 0 \rangle = \pm\frac{1}{\sqrt{6}} \big[ & |\bar{\zeta}\eta\tau\bar{\gamma}'| + |\zeta\bar{\eta}\tau\bar{\gamma}'| + |\zeta\eta\bar{\tau}\bar{\gamma}'| + \\ & + |\zeta\bar{\eta}\bar{\tau}\gamma'| + |\bar{\zeta}\eta\bar{\tau}\gamma'| + |\bar{\zeta}\bar{\eta}\tau\gamma'| \big] \end{aligned} \tag{A6c}$$

$$\begin{aligned} | \, {}^5E_g\gamma \breve{1} \rangle = \pm\frac{1}{2} \big[ & |\bar{\zeta}\bar{\eta}\bar{\tau}\gamma'| + |\zeta\bar{\eta}\bar{\tau}\bar{\gamma}'| + |\bar{\zeta}\eta\bar{\tau}\bar{\gamma}'| + \\ & + |\bar{\zeta}\bar{\eta}\tau\bar{\gamma}'| \big] \end{aligned} \tag{A6d}$$

$$| \, {}^5E_g\gamma \breve{2} \rangle = \pm|\bar{\zeta}\bar{\eta}\bar{\tau}\bar{\gamma}'| \tag{A6e}$$

Here the bases for the irreducible representation ${}^5E_g$ are $\gamma, \gamma' = u, v$ being $\gamma \neq \gamma'$ and the positive $(+)$ signs corresponding to $\gamma = u$ and the negative $(-)$ signs corresponding to $\gamma = v$.

### b. $D_{4h}$ point symmetry

When the cell is tetragonally distorted by a Jahn-Teller instability (reducing the symmetry to $D_{4h}$), the $T_{1g}$ representation is broken into a representation $A_{2g}$ with orbital symmetry $\gamma = \nu$, and $E_g$ with symmetries $\gamma = \kappa, \mu$. On the other hand, the reduction to $D_{4h}$ symmetry splits the $E_g$ representation into $A_{1g}$ with basis $\gamma = u$ and $B_{1g}$ with basis $\gamma = v$. Therefore, in $D_{4h}$ point symmetry the terms are split in the following representations (we show only the terms with maximum spin quantum number):

$$| \, {}^3A_{2g}\nu 1 \rangle = |\zeta\eta\tau\bar{\tau}| \tag{A7a}$$

$$| \, {}^3E_g\kappa 1 \rangle = |\zeta\eta\tau\bar{\zeta}| \tag{A7b}$$

$$| \, {}^3E_g\mu 1 \rangle = |\zeta\eta\tau\bar{\eta}| \tag{A7c}$$

$$| \, {}^5A_{1g}u 2 \rangle = |\zeta\eta\tau v| \tag{A7d}$$

$$| \, {}^5B_{1g}v 2 \rangle = |\zeta\eta\tau u| \tag{A7e}$$

We note that under a tetragonal elongated distortion the term ${}^5B_{1g}$ is lowest in energy. Although in the mono-electronic picture the fourth electron that drives the Jahn-Teller instability sits in an orbital with $u$ symmetry, the many-electron wave-function of the ${}^3B_{1g}$ term has $v$ symmetry. We also note that we introduce a global phase $-1$ in the term ${}^5B_{1g}$ to eliminate a minus sign.

### c. Spin-orbit coupling

Finally, we analyze how spin-orbit coupling splits the representations expressed in Eq. (A7). We remind that under spin-orbit coupling, wavefunctions are expressed as double group representations $|\Gamma\gamma\rangle$. In Eq. (A8) and Eq. (A9), double group representations (on the left side) are expressed in terms of the irreducible representations in $D_{4h}$ point symmetry (right side). First of all, note that the only term that splits under spin-orbit coupling is ${}^3E_g$. This can be understood by observing the reduced matrices in Eq. (7a) of the main text: only the reduced matrix $\mathbf{V}^{A_{2g}}$ has one non-zero diagonal element corresponding to this term. The spin part of ${}^3E_g$ cannot be described by a $j = 1$ representation in continuous rotation group because spins interact with the orbital space, which, in this case, is described by the $E_g$ representation in $D_{4h}$. In this symmetry, the continuous spin rotation group $D^{(S=1)}$ decomposes into $A_{2g} + E_g$ representations –which are *gerade*, since spinors are even under parity inversion [76]–, with representations in spherical basis $A_{2g}$ with $q = 0$, and $E_g$ with $q = \pm 1$. Therefore, we need to obtain the representation of the composite product $(A_{2g} + E_g) \otimes E_g = A_{2g} \otimes E_g + E_g \otimes E_g$. Thus, for $A_{2g} \otimes E_g$, i.e., when the orbital component $E_g$ couples to the $A_{2g}$ representation in the spin space, it generates two functions that transform as $E_g$ representations:

$$|E_g\kappa\rangle = |\,^3E_g\mu 0\rangle \tag{A8a}$$

$$|E_g\mu\rangle = -|\,^3E_g\kappa 0\rangle \tag{A8b}$$

On the other hand, the $E_g$ representation of the orbital part combines with the $E_g$ representation of the spin part, giving the following irreducible representations $E_g \otimes E_g = A_{1g} + A_{2g} + B_{1g} + B_{2g}$, which are expressed as follows.

$$|A_{1g}\rangle = -\frac{1}{2}\left[|\,^3E_g\kappa 1\rangle - |\,^3E_g\kappa\breve{1}\rangle - \imath|\,^3E_g\mu 1\rangle - \imath|\,^3E_g\mu\breve{1}\rangle\right] \tag{A9a}$$

$$|A_{2g}\rangle = \frac{\imath}{2}\left[|\,^3E_g\kappa 1\rangle + |\,^3E_g\kappa\breve{1}\rangle - \imath|\,^3E_g\mu 1\rangle + \imath|\,^3E_g\mu\breve{1}\rangle\right] \tag{A9b}$$

$$|B_{1g}\rangle = \frac{1}{2}\left[|\,^3E_g\kappa 1\rangle - |\,^3E_g\kappa\breve{1}\rangle + \imath|\,^3E_g\mu 1\rangle + \imath|\,^3E_g\mu\breve{1}\rangle\right] \tag{A9c}$$

$$|B_{2g}\rangle = -\frac{\imath}{2}\left[|\,^3E_g\kappa 1\rangle + |\,^3E_g\kappa\breve{1}\rangle + \imath|\,^3E_g\mu 1\rangle - \imath|\,^3E_g\mu\breve{1}\rangle\right] \tag{A9d}$$

Note that we used the Clebsch-Gordan coefficients displayed in Table II in Appendix C to obtain the expressions in Eq. (A8) and Eq. (A9). On the other hand, for some calculations it may be convenient to express Eq. (A8) and Eq. (A9) in spherical basis for the orbital angular momentum of the $^3E_g$ term, which contains components with quantum numbers $M_L = \pm 1$ labeled as $t_\pm$, and $t_0$. We can then rewrite the corresponding terms as:

$$|\,^3E_g t_\pm M\rangle = \mp\frac{1}{\sqrt{2}}\left[|\,^3E_g\kappa M\rangle \pm \imath|\,^3E_g\mu M\rangle\right] \tag{A10}$$

$$|A_{1g}\rangle = -\frac{1}{\sqrt{2}}\left[|\,^3E_g t_+\breve{1}\rangle + |\,^3E_g t_-1\rangle\right] \tag{A11a}$$

$$|A_{2g}\rangle = -\frac{\imath}{\sqrt{2}}\left[|\,^3E_g t_+\breve{1}\rangle - |\,^3E_g t_-1\rangle\right] \tag{A11b}$$

$$|E_g\kappa\rangle = \frac{\imath}{\sqrt{2}}\left[|\,^3E_g t_+0\rangle - |\,^3E_g t_-0\rangle\right] \tag{A11c}$$

$$|E_g\mu\rangle = -\frac{1}{\sqrt{2}}\left[|\,^3E_g t_+0\rangle + |\,^3E_g t_-0\rangle\right] \tag{A11d}$$

$$|B_{1g}\rangle = -\frac{1}{\sqrt{2}}\left[|\,^3E_g t_+1\rangle + |\,^3E_g t_-\breve{1}\rangle\right] \tag{A11e}$$

$$|B_{2g}\rangle = \frac{\imath}{\sqrt{2}}\left[|\,^3E_g t_+1\rangle - |\,^3E_g t_-\breve{1}\rangle\right] \tag{A11f}$$

Finally, we discuss how spin-orbit coupling splits the representation $^3E_g$ in $D_{4h}$ symmetry, see (Fig. 1b). For that purpose, we use the Wigner-Eckart theorem to compute the eigenenergies of the spin-orbit matrix elements. First, since $\langle E_g\kappa|E_g\gamma E_g\kappa\rangle = \langle E_g\mu|E_g\gamma E_g\mu\rangle = 0$ for $\gamma = \kappa, \mu, \nu$, the spin-orbit eigenenergies of the $E_g$ terms (see Eq. (A8)) are 0. On the other hand, by virtue of the expressions in Eq. (A11), the terms in Eq. (A9) can be expressed in the following way:

$$|\bar{\Gamma}\rangle = \sum_{\gamma,q} c_{\gamma q}|\,^3E_g\gamma q\rangle \tag{A12}$$

where $c_{\gamma q}$ can be obtained from Eq. (A11). The Clebsh-Gordan coefficients necessary to apply Wigner-Eckart are $\langle E_g\gamma|A_{2g}\nu E_g\gamma'\rangle = \left(\delta_\gamma^\kappa - \delta_\gamma^\mu\right)\left(1 - \delta_\gamma^{\gamma'}\right)$ and $\langle 1q|101q'\rangle = \delta_q^{q'} q/\sqrt{2}$, so that the matrix elements can be computed as:

$$\langle\bar{\Gamma}|\vec{L}\cdot\vec{S}|\bar{\Gamma}\rangle = \frac{\imath q}{\sqrt{2}}\sum_{\substack{\gamma,q\\ \gamma',q'}} c^*_{\gamma q}c_{\gamma' q'}\left(\delta_\gamma^\kappa - \delta_\gamma^\mu\right)\times \times\left(1 - \delta_\gamma^{\gamma'}\right)\delta_q^{q'} \tag{A13}$$

This gives matrix elements $\langle A_{1g}|\vec{L}\cdot\vec{S}|A_{1g}\rangle = \langle A_{2g}|\vec{L}\cdot\vec{S}|A_{2g}\rangle = -1/2$ and $\langle B_{1g}|\vec{L}\cdot\vec{S}|B_{1g}\rangle = \langle B_{2g}|\vec{L}\cdot\vec{S}|B_{2g}\rangle = 1/2$. As a result, spin-orbit coupling does not change the energy of the doubly degenerated $E_g$ spin-orbit term, while symmetric $(A_{1g}, A_{2g})$ and antisymmetric $(B_{1g}, B_{2g})$ representations split by $\mp\xi_{SO}/2$ with respect to the $E_g$ term (Fig. 1b). We also note that the accidental degeneracy of $(A_{1g}, A_{2g})$ and $(B_{1g}, B_{2g})$ terms may be eventually lifted if one considers developments beyond first-order relativistic contributions.

### Appendix B: Jahn-Teller Hamiltonian

Atoms or ions in a molecule or a unit cell have a position where the energy of the system is minimized. Sufficiently small deviations from these equilibrium positions can be described through a force constant:

$$K^{\Gamma\bar{\Gamma}} = \left(\frac{\partial^2 E^\Gamma}{\partial {Q^{\bar{\Gamma}}}^2}\right)_0 \tag{B1}$$

Here $Q^{\bar{\Gamma}}$ are the vibronic coordinates that transform under irreducible representation $\bar{\Gamma}$, which can be described in the frame of group theory being linear combinations of the displacements of the atoms in Cartesian coordinates $\Delta X_n, \Delta Y_n, \Delta Z_n$, and $E^\Gamma$ is the energy of the system which depends on the irreducible representation $\Gamma$ of the electronic wavefunction. In the presence of orbital degeneracy, the equilibrium positions change spontaneously, reducing the symmetry through the Jahn-Teller theorem [6]. This situation can be described by the addition of another term in the energy of the system that includes the potential energy of the nuclei in the field of the electrons in the state defined by the representation $\Gamma$ and basis $\gamma$, i.e., the adiabatic potential energy surface (APES) $\varepsilon_\gamma^\Gamma(\vec{Q})$ [9]:

$$\varepsilon(\vec{Q}) = \sum_{\Gamma,\bar{\Gamma}}\left[\frac{1}{2}K^{\Gamma\bar{\Gamma}}Q^2 + \varepsilon_\gamma^\Gamma(\vec{Q})\right] \tag{B2}$$

where $\varepsilon_\gamma^\Gamma(\vec{Q})$ is obtained by solving the secular equation for the vibronic coupling matrix operator $W$ which, to second order, is defined as:

$$W(r,Q) = \sum_{\Gamma\gamma}\left(\frac{\partial V}{\partial Q_\gamma^\Gamma}\right)_0 Q_\gamma^\Gamma + \\ + \frac{1}{2}\sum_{\Gamma'\gamma'\Gamma''\gamma''}\left(\frac{\partial^2 V}{\partial Q_{\gamma'}^{\Gamma'}\partial Q_{\gamma''}^{\Gamma''}}\right)_0 Q_{\gamma'}^{\Gamma'} Q_{\gamma''}^{\Gamma''} \tag{B3}$$

where $V$ refers to the electron-ion interaction potential. We can thus define first-order Eq. (B4a) and second-order Eq. (B4b) vibronic coupling terms as follows [9]:

$$X_\gamma^\Gamma = \left(\frac{\partial V}{\partial Q_\gamma^\Gamma}\right)_0 \tag{B4a}$$

$$X_{\gamma_1\gamma_2}^{\Gamma_1\Gamma_2} = \left(\frac{\partial^2 V}{\partial Q_{\gamma_1}^{\Gamma_1}\partial Q_{\gamma_2}^{\Gamma_2}}\right)_0 \tag{B4b}$$

These operators transform as the representation of the group corresponding to the lattice distortions [9], so for the computation of the matrix elements we can use the Wigner-Eckart theorem.

For the $E\otimes e$ problem, the following matrix elements can be derived using the functions defined in Eq. (A3):

$$F_E = \langle v|X_u^{E_g}|v\rangle \tag{B5}$$

$$G_E = \langle u|X_{vv}^{E_gE_g}|u\rangle \tag{B6}$$

By using the Wigner-Eckart theorem we can develop the corresponding Hamiltonian as:

$$\mathcal{H}_{JT}^{E\otimes e} = \frac{1}{2}K_E\rho^2\upsilon_0 + \\ \left[F_E\rho\cos\vartheta + G_E\rho^2\cos(2\vartheta)\right]\upsilon_z + \\ + \left[F_E\rho\sin\vartheta - G_E\rho^2\sin(2\vartheta)\right]\upsilon_x \tag{B7}$$

where $\upsilon_i$ are the Pauli matrices in the pseudospin space of $\{v,u\}$ and the vibronic coordinates have been normalized as $Q_2 = \rho\sin\vartheta$ and $Q_3 = \rho\cos\vartheta$. The eigenstates of this Hamiltonian are:

$$\varepsilon_E(\rho,\vartheta) = \frac{1}{2}K_E\rho^2 \pm \\ \pm\rho\sqrt{F_E^2 + G_E^2\rho^2 + 2F_EG_E\rho\cos(3\vartheta)} \tag{B8}$$

with the following eigenstates:

$$w_+ = \frac{1}{\sqrt{2}}\left(v\cos\frac{\Omega}{2} + u\sin\frac{\Omega}{2}\right) \tag{B9a}$$

$$w_- = \frac{1}{\sqrt{2}}\left(u\cos\frac{\Omega}{2} - v\sin\frac{\Omega}{2}\right) \tag{B9b}$$

The energy minima are found when $\vartheta = 2n\pi/3$, $n = 0,1,2$ –which corresponds to tetragonal elongations along the three axes– and $\rho = F_E/(K_E - 2G_E)$. In order to simplify the computation, the radial variable is normalized to $\rho = 1$. Around the tetragonal elongations, the parameter $\Omega$ in Eq. (B9), which is defined as:

$$\tan\Omega = \frac{F_E\sin\vartheta + |G_E|\sin(2\vartheta)}{F_E\cos\vartheta - |G_E|\cos(2\vartheta)} \tag{B10}$$

can be approximated as $\Omega \approx \vartheta$. In this situation, $w_-$ has a $u$-like symmetry and $w_+$ has a $v$-like symmetry. Since we consider tetragonal elongations with orthorhombic perturbations ($\vartheta = 2n\pi/3 \pm \delta\vartheta$), we can work with the following basis:

$$\check{v} = \frac{1}{\sqrt{2}}\left(v\cos\frac{\vartheta}{2} + u\sin\frac{\vartheta}{2}\right) \tag{B11a}$$

$$\check{u} = \frac{1}{\sqrt{2}}\left(u\cos\frac{\vartheta}{2} - v\sin\frac{\vartheta}{2}\right) \tag{B11b}$$

With this basis, and taking into account $\delta\vartheta$, the expression Eq. (B7) is transformed as:

$$\mathcal{H}_{JT}^{E\otimes e} = \frac{F_E + 2G_E}{2}\upsilon_0 + (F_E + G_E)\upsilon_z + \\ + (F_E - 2G_E)\delta\vartheta\upsilon_x \tag{B12}$$

As argued in the main text, in the $t_{2g}$ shell we only consider the $T\otimes e$ problem. For this problem we also neglect second order vibronic constants. The $T\otimes e$ vibronic constant is defined as:

$$F_T = \langle\tau|Q_u^{E_g}|\tau\rangle \tag{B13}$$

Using again the Wigner-Eckart theorem, the Hamiltonian of the $T\otimes e$ Jahn-Teller interaction is derived as a function of Gell-Mann matrices, $\lambda_k$, in the basis $\{\zeta,\eta,\tau\}$:

$$\mathcal{H}_{JT}^{T\otimes e} = \frac{1}{2}K_T\rho^2\lambda_0 - \frac{1}{2}F_T\rho\left[\sqrt{3}\lambda_8\cos\vartheta + \\ + \lambda_3\sin\vartheta\right] \tag{B14}$$

The energy minima correspond again to the tetragonal elongations with $\rho = F_T/K_T$. Since the nuclei motion is much slower than the electronic transitions, we can make the assumption that these minima are the same as the ones for the $E\otimes e$ problem, so we normalize again $\rho = 1$, so that $F_T = K_T$. These minima correspond to the basis defined before. We can generalize the expression Eq. (B14) to the local basis at each value $\vartheta_n$, denoted as $\{\check{\zeta},\check{\eta},\check{\tau}\}$, which is defined by rotations around the [111] axis in the orbital space, i.e., $\check{\tau} = \hat{R}_3^n(xyz)\tau$, where $\hat{R}_k^n(x...)$ defines a rotation of k-th order (angle $2\pi/k$) executed $n$ times along the axis defined by the coordinates in the parentheses. Then, the orthorhombic distortions in $T\otimes e$ are described by:

$$\mathcal{H}_{JT}^{T\otimes e} = \frac{1}{2}F_T\left[\lambda_0 - \sqrt{3}\lambda_8 - \delta\vartheta\lambda_3\right] \tag{B15}$$

The description so far is done with monoelectronic orbitals. We can generalize these results to many-electron wavefunctions. The vibronic constants are calculated using a one-body potential, so that for the non-diagonal matrix elements we only need to check the orbitals that are different (see also Appendix D). All the wavefunctions described by Eq. (A5) and Eq. (A6) are defined in the orbital part in terms of determinants of the type $|\zeta\eta\tau\gamma|$. Since there is just one different orbital in each Slater determinant, the off-diagonal elements are not modified. Then, for the first order vibronic constants, since the sums for $\langle t|X_u^{E_g}|t\rangle$ and $\langle t|X_v^{E_g}|t\rangle$ for $t=\zeta,\eta,\tau$ are null, the results for many-electron wavefunctions are the same as for the monoelectronic orbitals. We note that the same arguments apply for the $T\otimes e$ problem, since second order vibronic constants are neglected.

## Appendix C: Reduced matrix elements of the spin-orbit coupling operator

For the computation of the matrix elements of the spin-orbit coupling, we use the Wigner-Eckart theorem applied to the spin-orbit operator $V^{\Lambda}_{\lambda q}$ defined in Sec. II B, see also Ref. [1]. This operator transforms according to irreducible representations $\Lambda$ in the orbital space with basis $\lambda$ and $S^1_q$ corresponds to irreducible representations in the spin-rotation group. To calculate a given matrix element, we apply the Wigner-Eckart theorem as follows:

$$\begin{aligned}\langle \Gamma\gamma SM|V^{\Lambda}_{\lambda q}|\Gamma'\gamma' S'M'\rangle = \frac{(-1)^{1-g_\Lambda}}{\sqrt{g_\Gamma(2S+1)}}\times\\ \times\langle\Gamma S||\mathbf{V}^{\Lambda}||\Gamma' S'\rangle\langle\Gamma\gamma|\Lambda\lambda\Gamma'\gamma'\rangle\times\\ \times\langle SM|1qS'M'\rangle\end{aligned} \tag{C1}$$

where $|\Gamma\gamma SM\rangle$ and $|\Gamma'\gamma' S'M'\rangle$ correspond to wavefunctions that transform as irreducible representations $\Gamma$, $\Gamma'$ in bases $\gamma$, $\gamma'$ with spin $S$, $S'$ and spin quantum numbers $M$, $M'$, while $g_\Lambda$ and $g_\Gamma$ are the dimensionality of representations $\Lambda$ and $\Gamma$. The application of the Wigner-Eckart theorem requires the computation of the reduced matrices $\langle\Gamma S||\mathbf{V}^{\Lambda}||\Gamma' S'\rangle$. The latter have to be hermitic, which, as will be shown below, is guaranteed by the following expression [1]:

$$\begin{aligned}&\langle \Gamma'\gamma' S'M'|V^{\Lambda}_{\lambda q}|\Gamma\gamma SM\rangle\\ &\qquad= -(-1)^q\langle\Gamma\gamma SM|V^{\Lambda}_{\lambda\check{q}}|\Gamma'\gamma' S'M'\rangle\end{aligned} \tag{C2}$$

where, since $\vec{S}$ is expressed in spherical coordinates, we have $q=+1,0,-1$. One has to consider also the following relations between the Clebsh-Gordan coefficients:

$$\begin{aligned}\langle SM|1qS'M'\rangle = (-1)^{S-S'+q}\sqrt{\frac{2S+1}{2S'+1}}\times\\ \times\langle S'M'|1\check{q}SM\rangle\end{aligned} \tag{C3a}$$

$$\langle SM|1qS'M'\rangle = (-1)^{1+S'-S}\langle SM|S'M'1q\rangle \tag{C3b}$$

$$\langle\Gamma\gamma|\Lambda\lambda\Gamma'\gamma'\rangle = \sqrt{\frac{g_\Gamma}{g_{\Gamma'}}}\epsilon(\Gamma\Lambda\Gamma')\langle\Gamma'\gamma'|\Lambda\lambda\Gamma\gamma\rangle \tag{C3c}$$

$$\langle\Gamma\gamma|\Lambda\lambda\Gamma'\gamma'\rangle = \chi(\Gamma\Lambda\Gamma')\langle\Gamma\gamma|\Gamma'\gamma'\Lambda\lambda\rangle \tag{C3d}$$

The factors $\epsilon(\Gamma\Lambda\Gamma')$ and $\chi(\Gamma\Lambda\Gamma')$ depend on the phase convention [50]. We have fixed this convention by imposing $\langle\Gamma\gamma|A_{1g}u\Gamma\gamma\rangle = 1$ for any representation $\Gamma$ and basis $\gamma$. Assuming this convention, we have:

$$\epsilon(A_{1g}E_gE_g)=\epsilon(B_{1g}E_gE_g)=1 \tag{C4a}$$

$$\epsilon(A_{2g}E_gE_g)=\epsilon(B_{2g}E_gE_g)=-1 \tag{C4b}$$

$$\epsilon(A_{2g}A_{2g}A_{1g})=1 \tag{C4c}$$

$$\chi(E_gE_gA_{1g})=\chi(E_gE_gB_{1g})=1 \tag{C4d}$$

$$\chi(E_gE_gA_{2g})=\chi(E_gE_gB_{2g})=-1 \tag{C4e}$$

$$\chi(A_{2g}A_{2g}A_{1g})=1 \tag{C4f}$$

One can verify that Eq. (C3) and Eq. (C4) imply that $\langle\Gamma S||\mathbf{V}^{\Lambda}||\Gamma' S'\rangle = \langle\Gamma' S'||\mathbf{V}^{\Lambda}||\Gamma S\rangle$, which, as mentioned above, guarantees the expected hermiticity of the spin-orbit operator.

As shown in Sec. II B, the reduced matrices for the spin-orbit operator are expressed through irreducible representations $\mathbf{V}^{A_{2g}}$ and $\mathbf{V}^{E_g}$. To derive the matrix elements of these matrices, we use the Clebsh-Gordan coefficients expressed in Table II and the ladder operators defined as:

$$J_\pm|m\rangle = \sqrt{j(j+1)-m(m\pm1)}|m\pm1\rangle \tag{C5}$$

which are related to the spherical components of the angular momentum operators through:

$$J_\pm = \mp\frac{1}{\sqrt{2}}J_{\pm1} \tag{C6}$$

We first derive the matrix elements corresponding to the representation $\mathbf{V}^{A_{2g}}$. We remind that the reduced matrices are expressed in the basis $\{\,{}^3E_g,\ {}^3A_{2g},\ {}^5A_{1g},\ {}^5B_{1g}\}$ (see discussion in Sec. II B). We first note that the direct product $A_{2g}\otimes A_{2g}=A_{1g}$ implies that the spin-orbit operator in representation $A_{2g}$ has nonzero matrix elements connecting ${}^5A_{1g}$ and ${}^3A_{2g}$. Let us find such elements by applying the operators of angular and spin momenta to the wavefunctions of representation ${}^3A_{2g}$. We choose a spin-orbit operator in representation $q=1$ (corresponding to spin operator $s_+$) and $|\,{}^3A_{2g}\nu1\rangle$ expressed in terms of the corresponding Slater determinants (Eq. (A7a)). By applying the operators directly to the wavefunctions it follows that:

| $E$ | $E$ | $A_1$ $u$ | $A_2$ $\nu$ | $B_1$ $v$ | $B_2$ $\tau$ |
|---|---|---|---|---|---|
| $\kappa$ | $\kappa$ | $\frac{1}{\sqrt{2}}$ | 0 | $\frac{-1}{\sqrt{2}}$ | 0 |
| | $\mu$ | 0 | $\frac{-1}{\sqrt{2}}$ | 0 | $-\frac{1}{\sqrt{2}}$ |
| $\mu$ | $\kappa$ | 0 | $\frac{1}{\sqrt{2}}$ | 0 | $\frac{-1}{\sqrt{2}}$ |
| | $\mu$ | $\frac{1}{\sqrt{2}}$ | 0 | $\frac{1}{\sqrt{2}}$ | 0 |

TABLE II. Clebsh-Gordan coefficients for $E\otimes E$ in $D_{4h}$ group

$$\langle\, ^5A_{1g}u2|V^{A_{2g}}_{\nu 1}|\, ^3A_{2g}\nu 1\rangle = \\ = -\frac{1}{\sqrt{2}}\langle v|l_z|\tau\rangle\langle +\tfrac{1}{2}|s_+|-\tfrac{1}{2}\rangle = \imath\sqrt{2} \tag{C7}$$

On the other hand, by applying the Wigner-Eckart theorem (Eq. (C1)), we obtain:

$$\langle\, ^5A_{1g}u2|V^{A_{2g}}_{\nu 1}|\, ^3A_{2g}\nu 1\rangle = \\ = \frac{1}{\sqrt{5}}\langle\, ^5A_{1g}||\mathbf{V}^{A_{2g}}||\, ^3A_{2g}\rangle \tag{C8}$$

Combining Eq. (C7) and Eq. (C8) we obtain the reduced matrix element $\langle\, ^5A_{1g}||\mathbf{V}^{A_{2g}}||\, ^3A_{2g}\rangle = \imath\sqrt{10}$.

Next, we note that the direct product $E_g \otimes A_{2g} = E_g$ implies that $E_g$ wavefunctions can be connected through the spin-orbit operator only to wavefunctions of the same representation. This gives a diagonal element in the $\mathbf{V}^{A_{2g}}$ matrix. To find such element we apply, as before, the operators of angular and spin momenta to wavefunctions $^3E_g$ (Eq. (A7b) and Eq. (A7c)) and choose a representation $q = 0$, involving the spin operator $s_0$. The application of the operators to the wavefunctions gives:

$$\langle\, ^3E_g\kappa 1|V^{A_{2g}}_{\nu 0}|\, ^3E_g\mu 1\rangle = \\ = \langle\zeta|l_z|\eta\rangle\langle -\tfrac{1}{2}|s_0|-\tfrac{1}{2}\rangle = -\frac{\imath}{2} \tag{C9}$$

On the other hand, the application of the Wigner-Eckart theorem gives:

$$\langle\, ^3E_g\kappa 1|V^{A_{2g}}_{\nu 0}|\, ^3E_g\mu 1\rangle = \\ = -\frac{1}{\sqrt{12}}\langle\, ^3E_g||\mathbf{V}^{A_{2g}}||\, ^3E_g\rangle \tag{C10}$$

We therefore obtain $\langle\, ^3E_g||\mathbf{V}^{A_{2g}}||\, ^3E_g\rangle = \imath\sqrt{3}$. Using similar arguments, it can be shown that the rest of matrix elements of $\mathbf{V}^{A_{2g}}$ are zero, resulting in the reduced matrix described by Eq. (7a).

We derive now the matrix elements corresponding to the representation $\mathbf{V}^{E_g}$. We first note that the direct product $E_g \otimes E_g = A_{1g} \oplus A_{2g} \oplus B_{1g} \oplus B_{2g}$ means that the term $^3E_g$ can be connected by the spin-orbit operator to all other representations. As done for $\mathbf{V}^{A_{2g}}$, we combine the application of the operators of angular and spin momenta to the pertinent wavefunctions with the application of the Wigner-Eckart theorem. We obtain the following expressions:

$$\langle\, ^3A_{2g}\nu 1|V^{E_g}_{\kappa 0}|\, ^3E_g x1\rangle = \\ = \langle\tau|l_x|\zeta\rangle\langle -\tfrac{1}{2}|s_0|-\tfrac{1}{2}\rangle = \frac{\imath}{2} \\ = \frac{1}{\sqrt{12}}\langle\, ^3A_{2g}||\mathbf{V}^{E_g}||\, ^3E_g\rangle \tag{C11}$$

$$\langle\, ^5A_{1g}u2|V^{E_g}_{\kappa 1}|\, ^3E_g\kappa 1\rangle = \\ = \langle v|l_x|\zeta\rangle\langle +\tfrac{1}{2}|s_+|-\tfrac{1}{2}\rangle = -\frac{\imath}{\sqrt{2}} \\ = -\frac{1}{\sqrt{10}}\langle\, ^5A_{1g}||\mathbf{V}^{E_g}||\, ^3E_g\rangle \tag{C12}$$

$$\langle\, ^5B_{1g}v2|V^{E_g}_{\kappa 1}|\, ^3E_g\kappa 1\rangle = \\ = \langle u|l_x|\zeta\rangle\langle +\tfrac{1}{2}|s_+|-\tfrac{1}{2}\rangle = -\imath\sqrt{\frac{3}{2}} \\ = \frac{1}{\sqrt{10}}\langle\, ^5B_{1g}||\mathbf{V}^{E_g}||\, ^3E_g\rangle \tag{C13}$$

which allow us to obtain all the matrix elements for the $\mathbf{V}^{E_g}$ matrix as follows: $\langle\, ^3A_{2g}||\mathbf{V}^{E_g}||\, ^3E_g\rangle = \imath\sqrt{3}$, $\langle\, ^5A_{1g}||\mathbf{V}^{E_g}||\, ^3E_g\rangle = \imath\sqrt{5}$ and $\langle\, ^5B_{1g}||\mathbf{V}^{E_g}||\, ^3E_g\rangle = -\imath\sqrt{15}$.

These elements give the reduced matrix $\mathbf{V}^{E_g}$ expressed in Eq. (7b).

## Appendix D: One-body operators

In this work, Jahn-Teller and spin-orbit Hamiltonians contain one-body operators. In the same way, light-induced transfer requires also one-body operators in the electromagnetic Hamiltonian. In the following, we explain how one-body operators act in the formalism of the many-electron wavefunctions defined in Appendix A.

For that purpose, we recall that to comply with Pauli exclusion principle we need to define a multielectronic wavefunction $\Psi$ through the antisymmetrization operator $\mathcal{A}$ acting on the product of monoelectronic states occupied by electrons:

$$\Psi = \sqrt{N!}\mathcal{A}\prod_i \psi_i(i) = \frac{1}{\sqrt{N!}}\sum_{\sigma\in\mathcal{P}}(-1)^\sigma\prod_i \psi_{\sigma(i)}(i) \tag{D1}$$

where $\sigma$ is an element in the permutation group $\mathcal{P}$, $N$ is the number of fermions of the system and in $(-1)^\sigma$ represents the parity of the permutation. This results in the formation of Slater determinants.

We can define a one-body operator $O$ as the sum of operators $o_k$ acting over the k-th fermion as follows:

$$O = \sum_k o_k \tag{D2}$$

To find the matrix elements, we take into account the following properties of the antisymetrization operator:

- Applying $\mathcal{A}$ to a Slater determinant returns the

same Slater determinant, so that $\mathcal{A}^2 = \mathcal{A}$.

$$\begin{aligned}\mathcal{A}\Psi &= \frac{1}{\sqrt{N!}}\sum_{\sigma\in\mathcal{P}}(-1)^{\sigma}\mathcal{A}\prod_i \psi_{\sigma(i)}(i)\\ &= \frac{1}{(N!)^{3/2}}\sum_{\sigma,\tau\in\mathcal{P}}(-1)^{\sigma+\tau}\prod_i \psi_{\tau(\sigma(i))}(i)\\ &= \frac{1}{\sqrt{N!}}\sum_{\kappa\in\mathcal{P}}(-1)^{\kappa}\prod_i \psi_{\kappa(i)}(i)\end{aligned} \tag{D3}$$

We see that the composition of the two antisymmetrization operators defines another permutation in $\mathcal{P}$, $\kappa(i) = \tau(\sigma(i))$ with parity $\kappa = \tau + \sigma$, with $N!$ possible different compositions $\tau\sigma$ that return $\kappa$.

- Since $\mathcal{A}$ is a real operator $\mathcal{A}^\dagger = \mathcal{A}$.
- Since any one-body operator is even under permutations it always commute with $\mathcal{A}$, $[O, \mathcal{A}] = 0$.

Consequently, the matrix elements involving one-body operators between many-electron wavefunctions can be found as follows:

$$\begin{aligned}\langle\Phi|O|\Psi\rangle &= N!\langle\prod_j \phi_j(j)|\mathcal{A}^\dagger O\mathcal{A}|\prod_i \psi_i(i)\rangle = \sum_k\sum_{\sigma\in\mathcal{P}}(-1)^{\sigma}\langle\prod_j \phi_j(j)|o_k|\prod_i \psi_{\sigma(i)}(i)\rangle\\ &= \sum_k\sum_{\sigma\in\mathcal{P}}(-1)^{\sigma}\langle\phi_k(k)|o_k|\psi_{\sigma(k)}(k)\rangle\prod_i\langle\phi_i(i)|\psi_{\sigma(i)}(i)\rangle = \sum_k\langle\phi_k(k)|o_k|\psi_k(k)\rangle\prod_i\langle\phi_i(i)|\psi_i(i)\rangle\end{aligned} \tag{D4}$$

By orthogonality, the only permutation that does not vanish is the identity. Note that diagonal elements ($\Phi = \Psi$) do not vanish:

$$\langle\Psi|O|\Psi\rangle = \sum_k\langle\psi_k(k)|o_k|\psi_k(k)\rangle \tag{D5}$$

On the other hand, off-diagonal elements ($\Phi \neq \Psi$) can be nonzero only if the many-electron wavefunctions differ only by one one-body wavefunction. Otherwise, if they differ by more than one one-body wavefunction, the inner product vanishes.

$$\langle\Phi|O|\Psi\rangle = \langle\phi|o|\psi\rangle \tag{D6}$$